\documentclass[10pt,journal,compsoc]{IEEEtran}

\ifCLASSOPTIONcompsoc
  \usepackage[nocompress]{cite}
\else
  \usepackage{cite}
\fi
\ifCLASSINFOpdf
\else
\fi
\usepackage{graphicx}
\usepackage{booktabs}
\usepackage{hyperref}
\newcommand{\orcid}[1]{\href{https://orcid.org/#1}{\includegraphics[width=8pt]{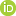}}}
\usepackage[table]{xcolor}
\usepackage{algorithm}
\usepackage{algorithmic}
\usepackage[utf8]{inputenc} 
\usepackage[T1]{fontenc}    
\usepackage{url}            
\usepackage{booktabs}       
\usepackage{amsfonts}       
\usepackage{nicefrac}       
\usepackage{microtype}      
\usepackage{xcolor}         
\usepackage{makecell}
\usepackage{adjustbox}
\usepackage{mathtools}
\usepackage{amsmath}
\usepackage{caption}
\usepackage{subcaption}
\usepackage{lipsum} 
\usepackage{dsfont} 
\usepackage{amssymb}
\usepackage{pifont}
\newcommand{\cmark}{\ding{51}}%
\newcommand{\xmark}{\ding{55}}%

\begin{document}

\title{\textls[-50]{HierSpeech++: Bridging the Gap between Semantic and Acoustic Representation of Speech by Hierarchical Variational Inference for Zero-shot Speech Synthesis}}
%
%
%
%

\author{Sang-Hoon Lee \orcid{0000-0002-8925-4474},
Ha-Yeong Choi \orcid{0000-0003-2390-7628},
Seung-Bin Kim \orcid{0000-0002-2287-9111},
and Seong-Whan Lee \orcid{0000-0002-6249-4996},~\IEEEmembership{Fellow,~IEEE}

\IEEEcompsocitemizethanks{

\IEEEcompsocthanksitem S.-H. Lee, H.-Y. Choi, S.-B. Kim, and S.-W. Lee are with the Department of Artificial Intelligence, Korea University, Seoul 02841, South Korea (e-mail: sh\_lee@korea.ac.kr; hayeong@korea.ac.kr; sb-kim@korea.ac.kr; sw.lee@korea.ac.kr)
\\
\textit{Corresponding author: Seong-Whan Lee}
\IEEEcompsocthanksitem This work was supported by Institute of Information \& communications Technology Planning \& Evaluation (IITP) grant funded by the Korea government (MSIT) (No. 2019-0-00079, Artificial Intelligence Graduate School Program (Korea University) and No. 2021-0-02068, Artificial Intelligence Innovation Hub).\protect\\
}

\thanks{Manuscript received November xx, 20xx; revised November xx, 20xx.}}

\markboth{Journal of \LaTeX\ Class Files,~Vol.~xx, No.~x, October~2023}%
{Shell \MakeLowercase{\textit{et al.}}: Bare Advanced Demo of IEEEtran.cls for IEEE Computer Society Journals}
%



\IEEEtitleabstractindextext{%
\begin{abstract}

Large language models (LLM)-based speech synthesis has been widely adopted in zero-shot speech synthesis. However, they require a large-scale data and possess the same limitations as previous autoregressive speech models, including slow inference speed and lack of robustness. This paper proposes HierSpeech++, a fast and strong zero-shot speech synthesizer for text-to-speech (TTS) and voice conversion (VC). We verified that hierarchical speech synthesis frameworks could significantly improve the robustness and expressiveness of the synthetic speech. Furthermore, we significantly improve the naturalness and speaker similarity of synthetic speech even in zero-shot speech synthesis scenarios. For text-to-speech, we adopt the text-to-vec framework, which generates a self-supervised speech representation and an F0 representation based on text representations and prosody prompts. Then, HierSpeech++ generates speech from the generated vector, F0, and voice prompt. We further introduce a high-efficient speech super-resolution framework from 16 kHz to 48 kHz. The experimental results demonstrated that the hierarchical variational autoencoder could be a strong zero-shot speech synthesizer given that it outperforms LLM-based and diffusion-based models. Moreover, we achieved the first human-level quality zero-shot speech synthesis. Audio samples and source code are available at \url{https://github.com/sh-lee-prml/HierSpeechpp}.

\end{abstract}
}

\maketitle

\IEEEdisplaynontitleabstractindextext

%
\IEEEpeerreviewmaketitle

\ifCLASSOPTIONcompsoc
\IEEEraisesectionheading{\section{Introduction}\label{sec:introduction}}
\else
\section{Introduction}
\label{sec:introduction}
\fi

The advent of large language models (LLM) have facilitated the widespread adoption of LLM-based models in speech synthesis and audio generation tasks. Conventional speech synthesis frameworks have advanced significantly driven by the integration of new features such as neural audio codecs for discrete speech or audio units. Although there is still a room for improvement in LLM-based speech models, these models possess four major limitations: 1) the first drawback is their auto-regressive generative manner, which has a slow inference speed and lack of robustness, resulting in repeating, skipping, and mispronunciation; 2) they are highly dependent on the pre-trained neural audio codec or discrete speech unit; 3) the audio quality of these models puts the clock back before the advent of the strong end-to-end speech synthesis framework proposed in \cite{kim2021conditional}; 4) they require a large-scale dataset to train the model.  

 VITS \cite{kim2021conditional} successfully introduced an end-to-end (E2E) speech synthesis framework by adopting a variational autoencoder (VAE) augmented with normalizing flow and adversarial training. Driven by the ability to generate high-quality waveform audio within a fully end-to-end training pipeline, the perceptual quality of synthetic speech is significantly better than that of two-stage speech synthesis models such as conventional text-to-speech (TTS) models and recent codec-based speech synthesis models. HierSpeech \cite{lee2022hierspeech} further improved the reconstruction quality by adopting a hierarchical conditional VAE using self-supervised speech representation. They proposed a novel TTS frameworks, which can train the model without any text transcripts by leveraging self-supervised speech representation and hierarchical VAE. However, the E2E models have limitations in terms of zero-shot voice cloning. Although E2E models can synthesize speech with high-quality audio, their synthetic speech still has a little speaker similarity in zero-shot voice cloning scenarios, and their training processes require high computational complexity.   

Meanwhile, diffusion-based speech synthesis models have also shown their strengths in terms of speaker adaptation. Diff-VC \cite{popov2022diffusionbased} introduced a conditional diffusion probabilistic model with a data-dependent prior for zero-shot voice conversion. Additionally, the effectiveness of diffusion models for speaker adaptation, including DDDM-VC \cite{choi2023dddm}, Diff-hierVC \cite{choi23d_interspeech}, and UnitSpeech \cite{kim2023unitspeech}, has been proven. Although diffusion models exhibit a good adaptation performance, they have several limitations: 1) they have a slow inference speed with their iterative generation processes, and 2) they are vulnerable to noisy data for speaker adaptation. With noisy speech prompts, they may generate much more noisy speech mainly due to the powerful adaptation performance, and this may further results in the degradation of the perceptual audio quality. 3) Although diffusion models have shown strong generative performance, they still possess lower audio quality owing the train-inference mismatch of the two-stage generation between the ground-truth and generated Mel-spectrogram.  


In this paper, we present HierSpeech++, a fast and strong zero-shot speech synthesis model that uses a hierarchical speech synthesis framework. By adopting the E2E speech synthesis frameworks to take the advantage of high-quality waveform generation, we solved the limitation of style adaptation by adopting a self-supervised speech representation as a semantic speech representation and bridging the gap between semantic and acoustic representation hierarchically. We propose a novel speech synthesis framework consisting of a hierarchical speech synthesizer, text-to-vec (TTV), and speech super-resolution (SpeechSR).

Based on HierVST \cite{lee23i_interspeech}, we introduce an improved hierarchical speech synthesizer using a hierarchical conditional VAE. To improve audio quality beyond perceptual quality, we adopt a dual-audio acoustic encoder in order to enhance the acoustic posterior and utilize a BigVGAN-based hierarchical adaptive generator with conditional and unconditional generation for better out-of-distribution generalization (zero-shot voice cloning). In addition, we adopt a source-filter theory-based multi-path semantic encoder to disentangle speech components and enhance the semantic prior for speaker-agnostic and speaker-related semantic information. By using a hierarchical VAE, we can connect and learn these representations hierarchically and infer the waveform audio by progressively adapting to the target voice style. For better adaptation and train-inference mismatch reduction, we introduce bidirectional normalizing flow Transformer networks using AdaLN-Zero. Without a text-speech paired dataset, we can simply scale-up the dataset to train a hierarchical speech synthesizer for zero-shot voice cloning.

For text-to-speech, we introduce a TTV that can generate a semantic representation and an F0 from text sequences. Owing to the semantic information that is extracted from self-supervised learning, we can transfer prosody information that is irrelevant to voice style. By connecting the TTV and a hierarchical speech synthesizer, we can synthesize high-quality speech from text by hierarchically adapting the prosody and voice style even in zero-shot scenarios. We also propose a simple speech super-resolution framework to upsample high-resolution waveform audio from 16 kHz to 48 kHz. This can facilitate data accessibility for scaling up datasets in that we can utilize low-resolution speech data such as the automatic speech recognition (ASR) dataset to train the speech synthesizer and TTV models. 

The main contributions of this study are as follows:   
\begin{itemize}
\item For fast and strong zero-shot speech synthesis, we presented HierSpeech++, a novel fully-parallel hierarchical speech synthesis framework. 
\item Prosody and voice style can be transferred and controlled using a hierarchical speech synthesis framework.
\item We also present SpeechSR which can upsample waveform audio from 16 kHz to 48 kHz for high-resolution speech synthesis and data scalability.
\item HierSpeech++ achieved the first human-level quality for zero-shot text-to-speech and voice conversion tasks.
\item Audio samples and source code are available at \url{https://sh-lee-prml.github.io/HierSpeechpp-demo/}
\end{itemize}

\section{Related Work}

\subsection{Neural Codec Language Models}
Conventional sequence-to-sequence auto-regressive TTS models, such as Tacotron \cite{wang17n_interspeech}, have successfully paved the way for speech synthesis technologies. TransformerTTS \cite{li2019neural} first adopted a Transformer network for TTS, and VTN \cite{9306875} also utilizes a Transformer network for VC. However, these auto-regressive models suffer from a slow inference speed, in addition to a lack of robustness owing to challenges in aligning text and acoustic representations and the difficulty in predicting a continuous acoustic representation. Recently, neural audio codec model \cite{zeghidour2021soundstream,dfossez2023high} have replaced conventional acoustic representations with a high-compressed audio codec, which can reproduce the original waveform audio. Vall-E \cite{wang2023neural} was the first neural codec language model for speech synthesis utilizing a discrete audio unit and language models. By scaling up the dataset to 60,000 h, Vall-E could perform in-context learning using a neural audio codec. However, it possessed the same limitations as auto-regressive TTS models, such as a slow inference speed and a lack of robustness. Furthermore they have a high-dependency of their pre-trained neural audio codec, resulting in low-quality audio. To overcome this limitation, high-quality neural audio codec models, such as HiFi-Codec \cite{yang2023hifi} and DAC \cite{kumar2023highfidelity}, have been investigated. Furthermore, SPEAR-TTS \cite{kharitonov2023speak} and Make-A-Voice \cite{huang2023make} introduced a hierarchical speech synthesis framework from semantic to acoustic token to reduce the gap between text and speech. Moreover, to reduce inference speed and improve the robustness of auto-regressive methods, SoundStorm \cite{borsos2023soundstorm} proposed parallel audio generation methods that generate the token of a neural audio codec. UniAudio \cite{yang2023uniaudio} presented a multi-scale Transformer architecture to reduce the computational complexity of long audio sequences.  

\subsection{Non-autoregressive Models}

For fast and robust speech synthesis, FastSpeech \cite{ren2019fastspeech} introduced a duration predictor to synthesize speech in parallel, and they significantly improved the robustness of speech synthesis by addressing the limitations of auto-regressive models such as repeating and skipping. To reduce the one-to-many mapping problem in non-autoregressive speech synthesis, FastSpeech 2 \cite{ren2020fastspeech} adopted a variance adaptor that can reflect pitch and energy information. However, these models require an external duration extractor to align the text and speech. Glow-TTS \cite{kim2020glow} introduces a monotonic alignment search and normalizing flow to learn text-speech alignment and train the TTS model simultaneously. They add a blank token interspersed between phoneme tokens to increase robustness. VITS \cite{kim2021conditional} combined the TTS model and a neural vocoder using VAE for end-to-end TTS frameworks with the aim of improving the quality of synthetic speech. NaturalSpeech \cite{tan2022naturalspeech} achieved human-level quality in a single speaker TTS by introducing a bidirectional normalizing flow and adopting a differentiable duration modeling and phoneme pre-training. Moreover, HierSpeech \cite{lee2022hierspeech} leveraged a self-supervised speech representation in end-to-end speech synthesis, which significantly reduced the information gap between text and speech, thus addressing speech mispronunciations. In addition, HierVST \cite{lee23i_interspeech} utilized a hierarchical VAE for zero-shot voice style transfer, and which significantly improved the voice style transfer performance in end-to-end speech synthesis models without any labels. ZS-TTS \cite{kim22c_interspeech}, WavthruVec \cite{siuzdak2022wavthruvec} and VQTTS \cite{du2022vqtts} utilized a self-supervised speech representation as an intermediate acoustic representation for robust speech synthesis. NANSY++ \cite{choi2022nansy++} introduced a unified speech synthesizer for various voice applications such as TTS, VC, singing voice synthesis, and voice control. Some studies \cite{jiang2023mega2} have combined a parallel TTS with LLM-based prosody modeling for expressive speech synthesis. 

\subsection{Diffusion Models}
Diffusion models have also demonstrated their powerful generative performances in speech synthesis. Grad-TTS \cite{popov2021grad} first introduced a score-based decoder to generate a Mel-spectrogram, and Diff-VC demonstrated the high-adaptation performance of diffusion models in zero-shot voice conversion scenarios. DiffSinger achieved a state-of-the-art performance in SVS task by generating a high-quality singing voice with a powerful adaptation performance. DDDM-VC \cite{choi2023dddm} significantly improved speech representation disentangle \cite{6472238} and voice conversion performance by a disentnalged denoising diffusion model and prior mixup. Diff-HierVC \cite{choi23d_interspeech} introduced a hierarchical voice style transfer frameworks that generates pitch contour and voice hierarchically based on diffusion models. Guided-TTS \cite{kim2022guided} and Guided-TTS 2 \cite{kim2022guided2} have also shown good speaker adaptation performance for TTS. UnitSpeech \cite{kim2023unitspeech} introduced a unit-based speech synthesis with diffusion models. Furthermore, recent studies utilized a diffusion model in latent representation. Naturalspeech 2 \cite{shen2023naturalspeech} and HiddenSinger \cite{hwang2023hiddensinger} utilized the acoustic representation of an audio autoencoder as a latent representation, and developed a conditional latent diffusion model for speech or singing voice synthesis. StyleTTS 2 \cite{li2023styletts} proposed a style latent diffusion for style adaptation. Although all the above models have shown powerful adaptation performance, they have a slow inference speed for their iterative generation manner. To reduce the inference speed, CoMoSpeech \cite{ye2023comospeech} and Multi-GradSpeech \cite{xue2023multi} adopted a consistency model for a diffusion-based TTS model. Recently, VoiceBox \cite{le2023voicebox} and P-Flow \cite{kim2023pflow} utilized flow matching with optimal transport for fast sampling. However, these models still have a training-inference mismatch problem that arises from two-stage speech synthesis frameworks and they are vulnerable to noisy target voice prompt.   
 
\subsection{Zero-shot Voice Cloning}
Zero-shot learning \cite{8413121} for voice cloning is a task to synthesize speech with a novel speaker, which has not been previously observed during training. A majority of the studies on voice cloning \cite{jia2018transfer,10073591} focused on cloning the voice styles, such as timbre and environment, and speaking styles, such as prosody and pronunciation. \cite{skerry2018towards} presented a reference encoder for prosody modeling, and GST \cite{wang2018style} utilized a learnable token for style modeling from the reference speech or manually control. \cite{lee2019robust} proposed a fine-grained prosody control for expressive speech synthesis from reference speech. Multi-SpectroGAN \cite{lee2021multi} utilized adversarial feedback and a mixup strategy for an expressive and diverse zero-shot TTS.  Meta-StyleSpeech \cite{min2021meta} introduced meta-learning for style modeling, and GenerSpeech \cite{huang2022generspeech} utilized a mix-style layer normalization for better generalization on the out-of-domain style transfer. PVAE-TTS \cite{lee2022pvae} utilized a progressive style adaptation for high-quality zero-shot TTS. AdaSpeech \cite{chen2021adaspeech} introduced adaptive layer normalization for adaptive speech synthesis. YourTTS \cite{casanova2022yourtts} trained the VITS \cite{kim2021conditional} with a speaker encoder and Grad-StyleSpeech \cite{kang2023grad} utilized a style-conditioned prior on a score-based Mel-spectrogram decoder for better adaptive TTS. Built upon VQTTS \cite{du2022vqtts}, TN-VQTTS \cite{10229489} introduce a timbre-normalized vector-quantized acoustic features for speaking style and timbre transfer. Meanwhile, there are text prompt-based style generation models which can describe a speaking or voice style from text descriptions \cite{yang2023instructtts,guo2023prompttts,leng2023prompttts}.

\begin{figure*}[t]
    \centering
    {\includegraphics[width=1\textwidth]{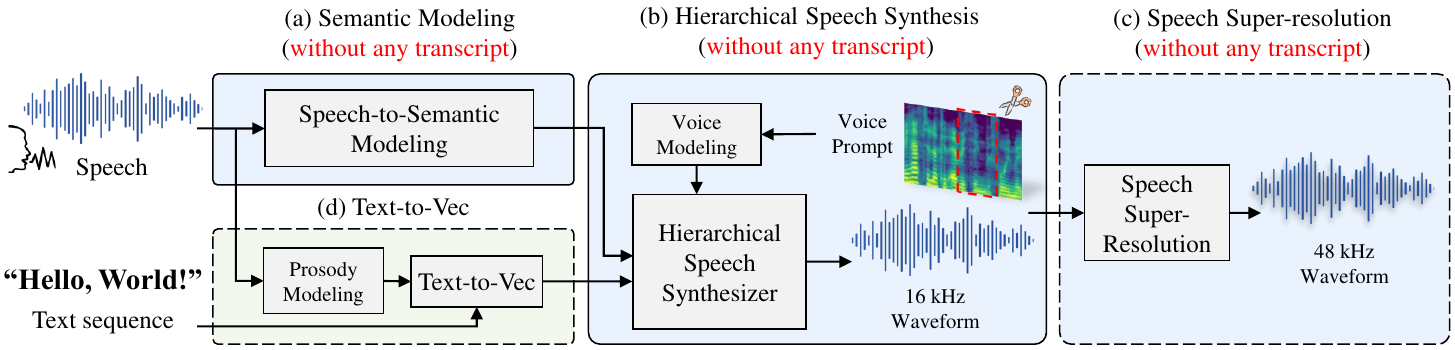}}
    \caption{Hierarchical speech synthesis pipeline of HierSpeech++. (a) For speech-to-semantic modeling, we use MMS which is a pre-trained Wav2Vec 2.0 with 1,406 language speech data. (b) Hierarchical speech synthesizer generates a 16 kHz waveform audio from semantic representation and voice prompt. (c) SpeechSR upsamples a 16 kHz waveform audio to 48 kHz. (d) For TTS, text-to-vec system generates a semantic representation from text sequence and prosody prompt. }
    \label{fg1}
\end{figure*}

\section{HierSpeech++}
In this study, we propose HierSpeech++, a human-level zero-shot speech synthesis model in terms of naturalness and voice similarity. We present a novel and efficient hierarchical speech synthesis framework that consists of a hierarchical speech synthesizer, text-to-vec (TTV), and speech super-resolution (SpeechSR) as illustrated in Fig \ref{fg1}. This framework facilitates the training of each model with scale-up in that we can simply utilize large-scale low-resolution speech data for voice cloning. The details are described in the following subsections.

\subsection{Speech Representations}

We utilize downsampled audio at 16 kHz for speech synthesizer because the human voice frequency band ranges under 4 kHz, and it is necessary to use at least twice the highest component of the voice frequency for the reconstruction of the voice signal. Furthermore, a low-resolution ASR dataset can be used to train the speech synthesizer. For better perceptual quality, we upsample the audio from 16 kHz to 48 kHz for a post-processing using SpeechSR. For acoustic and semantic representations, we utilize a low-resolution representation of 50 Hz for efficient training.

\subsubsection{Acoustic Representation}
For acoustic representation, conventional TTS systems utilize the Mel-spectrogram as an intermediate acoustic feature, which is transformed from the waveform using a short-time Fourier transform (STFT). Recently, a neural audio codec was investigated in the TTS model, wherein the Mel-spectrogram was replaced with a trained audio codec that can be decoded to waveform signals. However, acoustic features comprise various attributes, including semantic information, such as pronunciation and context, as well as voice information, such as timbre, intonation, and recording environment. In this regard, it is difficult to directly infer these rich representations from text by exacerbating the one-to-many mapping problem, and this may result in mispronunciation, over-smoothed speech, and a lack of similarity. To reduce the aforementioned problems, HierSpeech \cite{lee2022hierspeech} adopted a self-supervised speech representation as an additional semantic representation to bridge the gap between text and acoustic features as described in the following subsection.

\subsubsection{Semantic Representation}
We utilize Wav2Vec 2.0 \cite{baevski2020wav2vec} to extract a continuous semantic representation from a waveform without any label. In recent years, many researches have adopted a self-supervised speech representation for semantic representation given that the representation from the middle layer of these models contains linguistic information learned in a self-supervised manner. Meanwhile, phonetic posteriorgrams (PPG) that are extracted from ASR models \cite{9420297} or phoneme information \cite{9729483} could be a good alternative for a rich resourced mono-lingual model. However, this decreases the expressiveness and robustness of the zero-shot voice cloning and multi-lingual speech synthesis scenarios. Unlike HierSpeech \cite{lee2022hierspeech} which used XLS-R \cite{babu22_interspeech}, we utilize a massively multilingual speech (MMS) \cite{pratap2023scaling} which is a pre-trained Wav2Vec 2.0 with a massive scale. MMS was trained with 1,406 language speech data and it was observed that MMS performed better than XLS-R in many down-stream tasks. For zero-shot cross-lingual speech synthesis, we extract the semantic representation from the middle layer of the MMS.  

\subsubsection{Style Representation}
We use two global style representations for the prosody and voice style. For prosody style representation, we extract a global prosody representation from the Mel-spectrogram of the reference prosody prompt, which is conditioned to the TTV model. For voice style representation, we also extract a global prosody representation from the Mel-spectrogram of the reference voice prompt, and this is utilized for conditioning the hierarchical speech synthesizer. As we disentangle the semantic and acoustic modeling, we can transfer the prosody and voice styles in TTV and the hierarchical speech synthesizer separately. Although we extract both representations from the same Mel-spectrogram, each representation is trained to reflect each characteristic because each target representation contains semantic and acoustic information respectively. 


\begin{figure*}[ht]
    \centering
    \begin{subfigure}{.75\textwidth}
        \centering
        \includegraphics[width=1.\linewidth]{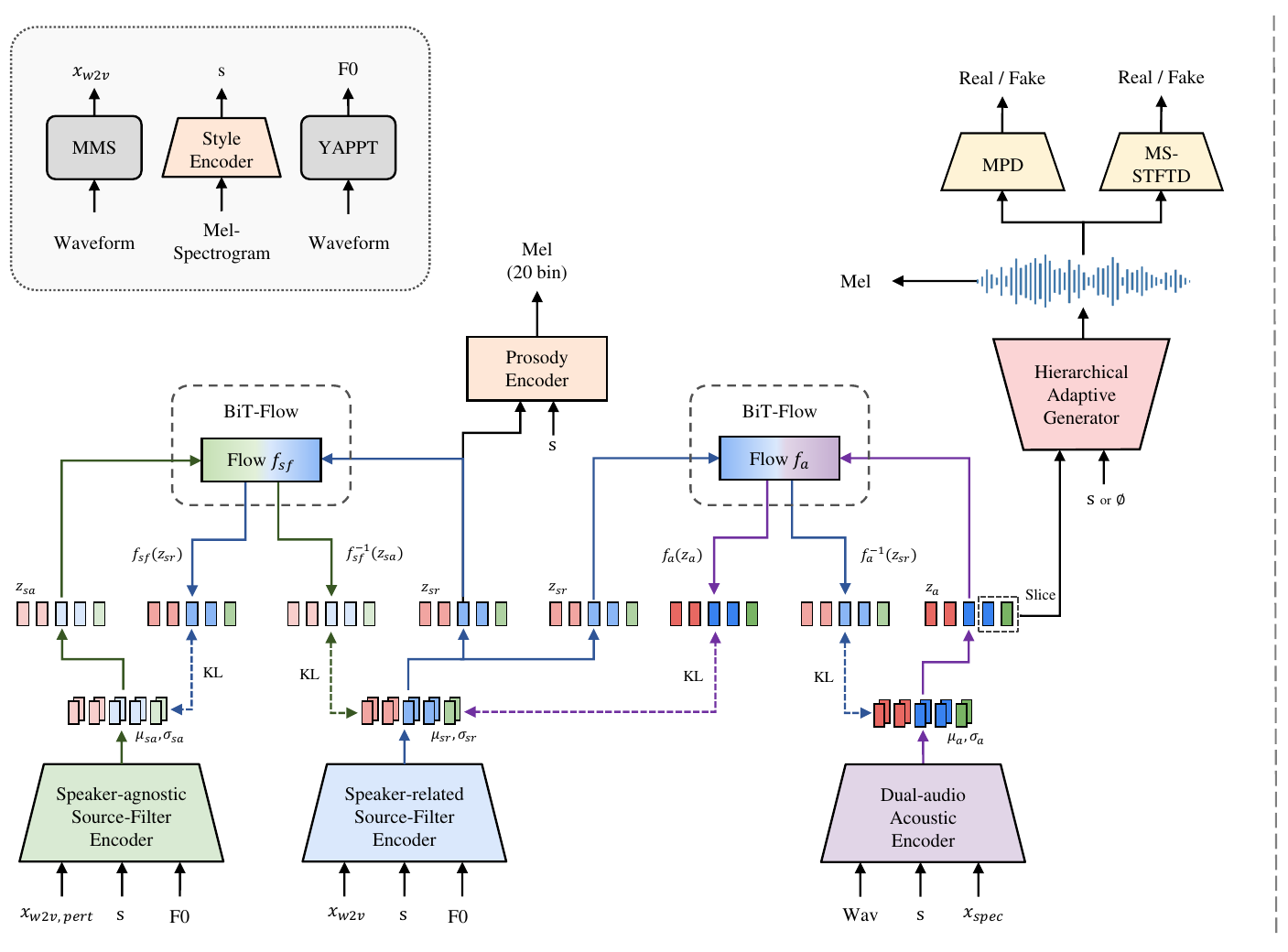} 
        \caption{Training}
        \label{fg2:train}
    \end{subfigure}\hfill 
    \begin{subfigure}{.25\textwidth}
        \centering
        \includegraphics[width=1.\linewidth]{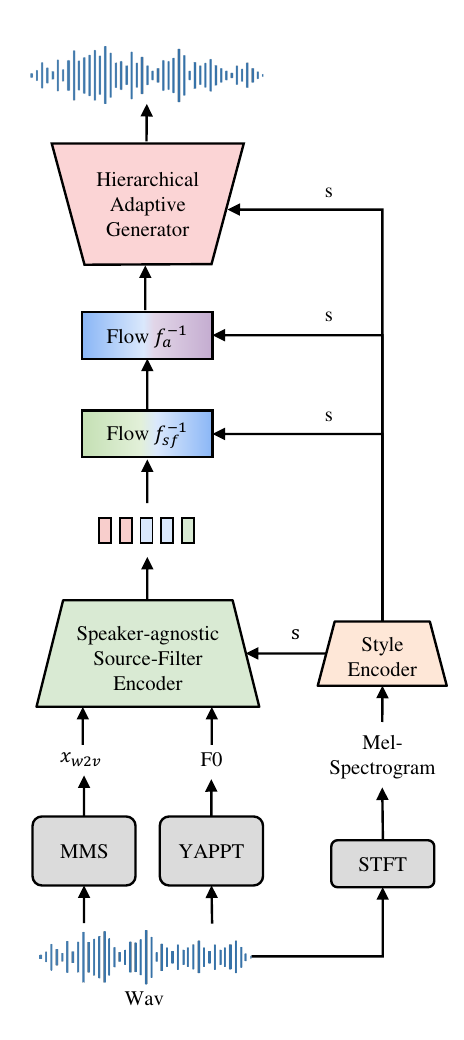} 
        \caption{Inference}
        \label{fg2:infer}
    \end{subfigure}\hfill   
    \caption{Hierarchical speech synthesizer of HierSpeech++.}
    \label{fg2}
\end{figure*}

\begin{figure*}[t]
    \centering
    {\includegraphics[width=1\textwidth]{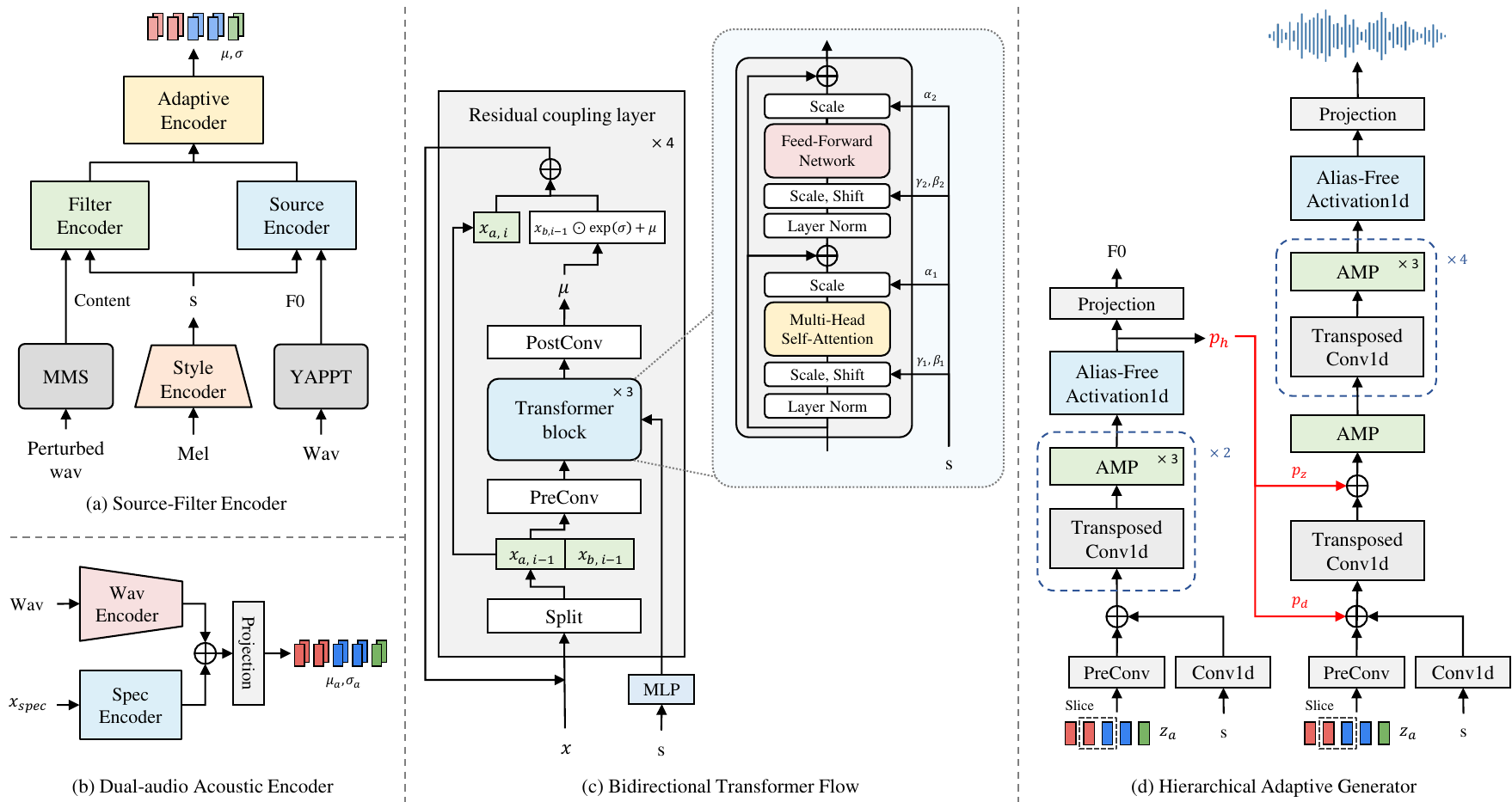}}
    \caption{Architecture details for hierarchical speech synthesizer. For inference, we can generate a waveform from semantic representation, F0, and voice prompt. The semantic representation and F0 can be extracted from source speech for voice conversion, and can generated by TTV for Text-to-speech.}
    \label{fg3}
\end{figure*}

\subsection{Hierarchical Speech Synthesizer}

We propse a hierarchical speech synthesizer as the backbone speech synthesizer for HierSpeech++. We can train this module using only speech data without any labels such as speaker id and text transcripts. Building on HierVST, we significantly improved the model for human-level audio quality and zero-shot style transfer, as illustrated in Fig \ref{fg2}.

\subsubsection{Dual-audio Acoustic Encoder}
Although previous models successfully increased the acoustic capacity by replacing the Mel-spectrogram with a linear spectrogram, the autoencoder that is trained by minimizing a KL divergence has shown low reconstruction quality in terms of PESQ, Mel-spectrogram distance, pitch periodity, and voice/unvoice F1 score. This may reduce a perceptual quality in zero-shot voice cloning. To address the limitations of using linear spectrogram and improve the perceptual quality, we introduce a dual-audio acoustic encoder designed to capture a more comprehensive and richer acoustic representation. We add a waveform encoder for distilling information from the raw waveform audio and concatenate a representation from waveform audio with a representation from a linear spectrogram. Finally, the acoustic representation is projected from a concatenated representation.  

\subsubsection{Source-filter Multi-path Semantic Encoder}
Following HierVST, we also utilize a multi-path self-supervised speech representation for speaker-related and speaker-agnostic semantic representations. Each representation is used as a prior for hierarchical style adaptation. We extract semantic representations from the middle layer of MMS to obtain linguistic information\footnote{Specifically, we utilize the representation from the seventh layer of MMS.}. We also utilize a fundamental frequency (F0) to improve speech disentanglement for enhanced prior, which enables a manual control of the pitch contour. For speaker-agnostic semantic representation, we utilize a speech perturbation to remove speaker-related information from the self-supervised speech representation. For speaker-related semantic representation, we do not use speech perturbation to reduce the gap between semantic and acoustic representation. 

\subsubsection{Hierarchical Variational Autoencoder}
We adopt the structure of HierSpeech \cite{lee2022hierspeech} wherein we replaced the text encoder with a linguistic restorer. We utilize the perturbed linguistic representation $x_{w2v,pert}$ as conditional information $c$ to hierarchically generate waveform audio. In addition we use an enhanced linguistic representation from the self-supervised representation of the original waveform, which is not perturbed. Moreover, the raw waveform audio is reconstructed from the acoustic representation which is extracted using a linear spectrogram and waveform audio during training. To connect the acoustic and multi-path linguistic representations, we utilize hierarchical variation inference. The optimization objective of a hierarchical speech synthesizer can be defined as follows:
\begin{equation}
\begin{split}
    \log{p_{\theta}}(x|c)&\geq \mathbb{E}_{q_{\phi}(z|x)}\Big[\log p_{\theta_d}(x|z_a)\\&-\log \frac{q_{\phi_{a}}(z_a|x)}{p_{\theta_a}(z_a|z_{sr})} -\log \frac{q_{\phi_{sr}}(z_{sr}|x_{w2v},F0)}{p_{\theta_{sr}}(z_{sr}|c)} \Big]
    \end{split}
\end{equation}
where $q_{\phi_a}(z_a|x)$ and $q_{\phi_{sr}}(z_{sr}|x_{w2v},F0)$ are the approximate posteriors for the acoustic and linguistic representations respectively. Here, $p_{\theta_{sr}}(z_{sr}|c)$ represents a prior distribution of linguistic latent variables $z_{sr}$ given condition $c$, $p_{\theta_a}(z_a|z_{sr})$ denotes a prior distribution on acoustic latent variables $z_a$, and $p_{\theta_d}(x|z_a)$ is the likelihood function represented by a HAG that produces data $x$ given latent variables $z_a$. Condition $c$ consists of a perturbed representation of $x_{w2v,pert}$ and a log-scale fundamental frequency $F_0$. In addition, we use a normalizing flow to improve the expressiveness of each linguistic representation. For reconstruction loss, we use multiple reconstruction terms of a HAG with adversarial training as described in the following subsection.

\subsubsection{Hierarchical Adaptive Generator}
For semantic-to-waveform generation, we introduce the HAG $G$ which comprises of source generator $G_s$ and waveform generator $G_w$ as illustrated in Fig \ref{fg3}. The generated representations including acoustic representation $z_a$ and style representation $s$ are fed to $G_s$, and $G_s$ generates the refined pitch representation $p_h$ and auxiliary F0 predictor is used to enforce the F0 information on $p_h$ as follows:

\begin{equation}
\label{Pitch_loss}
   L_{pitch} = \lVert p_x-G_s(z_a, s)\rVert_1,
\end{equation}
where $p_x$ is the ground-truth F0. Subsequently, $G_w$ synthesizes the waveform audio from $z_a, p_h, s$ hierarchically, and we use the reconstruction loss between the GT and generated Mel-spectrogram transformed from waveform audio using STFT with Mel-filter $\psi$ as follows:
\begin{equation}
\label{Mel_loss}
   L_{STFT} = \lVert \psi(x)-\psi(G_w(z_a, p_h, s))\rVert_1.
\end{equation}
For better perceptual audio quality, we utilize adversarial training \cite{lee2021multi,chung21_interspeech}. We adopt the multi-period discriminator (MPD) \cite{kong2020hifi} and the multi-scale STFT discriminator (MS-STFTD) \cite{defossez2022high} which can reflect the characteristics of the real and imaginary components from a complex-valued STFT as: 
\begin{equation}
  \mathcal{L}_{adv}(D) = \mathbb{E}_{(x,z_a)}\Big[(D(x)-1)^2 + D(G(z_a, s))^2 \Big],
\end{equation}
\begin{equation}
  \mathcal{L}_{adv}(\phi_a, \theta_d) = \mathbb{E}_{(z_a)}\Big[(D(G(z_a, s))-1)^2 \Big]
\end{equation}

\subsubsection{Bidirectional Transformer Flow}

Previously, VITS utilized a normalizing flow to improve the expressiveness of the prior distribution, which possessed the ability to bridge the posterior and prior distributions well, such that the audio quality was significantly improved. However, there is a train-inference mismatch problem and limitations on speaker adaptation in this framework. Therefore, we propose BiT-Flow, which is a bidirectional Transformer-based normalizing flow. First, we adopt the bidirectional normalizing flow proposed by NaturalSpeech \cite{shen2023naturalspeech}. We also replace WaveNet-based adaptive networks with convolutional feedforward-based Transformer networks \cite{9716741} to capture a large context in the latent representation. Unlike VITS2, we utilize AdaLN-Zero in Transformer block \cite{peebles2023scalable} and train the blocks bidirectionally with dropout. The details are illustrated in Fig \ref{fg3} For efficient training, we sliced audio sequence during training the model so we do not utilize a positional embedding in Transformer networks. 

\subsubsection{Prosody Distillation}
We utilize prosody distillation to enhance the linguistic representation $z_{sr}$ from the speaker-related source-filter encoder. $z_{sr}$ is fed to the prosody decoder to reconstruct the first 20 bins of the Mel-spectrogram containing the prosody information. By conditioning the voice style representation, we enforce $z_{sr}$ to learn speaker-related prosody to enhance linguistic information. We train the model with prosody loss $\mathcal{L}_{prosody}$ which minimizes the $l1$ distance between the 20 bins of the GT and the reconstructed Mel-spectrogram.

\subsubsection{Unconditional Generation}
Following \cite{lee23i_interspeech}, we utilize unconditional generation in a hierarchical adaptive generator for progressive speaker adaptation. The use of unconditional generation makes the acoustic representations adopt speaker characteristics, thus improving the speaker adaptation performance in the entire model. We simply replace the voice style representation $s$ with the null style embedding $\varnothing$ with a 10\% chance during training. In the inference stage, we utilize only the target voice style representation for conditional generation.   

\subsection{Text-to-Vec}
\begin{figure}[t]
    \centering
    {\includegraphics[width=1\columnwidth]{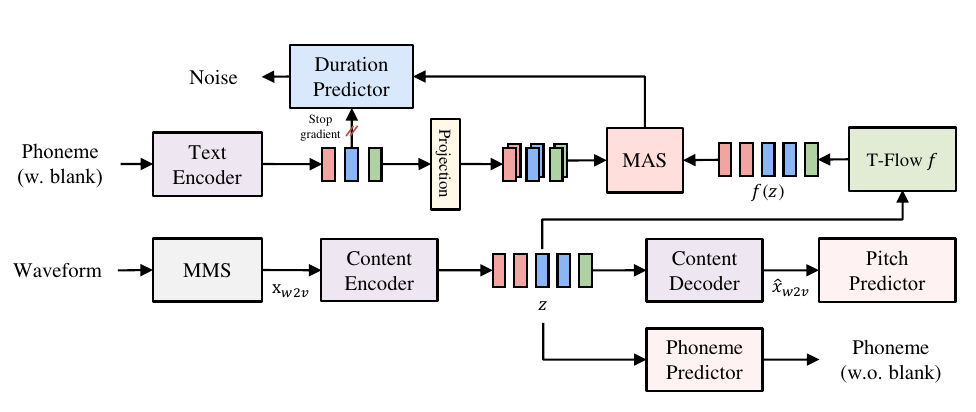}}
    \caption{Text-to-Vec}
    \label{fg4}
\end{figure}

For TTS, we introduce a text-to-vec (TTV) model that generates a semantic representation and F0 from a text sequence. Following VITS \cite{kim2021conditional}, we utilize a variational autoencoder and a monotonic alignment search (MAS) to align the text and speech internally, as shown in Fig \ref{fg4}. We replace the linear spectrogram with a self-supervised speech representation for the input of posterior encoder, and we reconstruct the same self-supervised speech representation for the output of TTV. Furthermore, we predict a F0 with four$\times$ larger resolutions than the self-supervised speech representation. We use a text sequence and prosody prompt as conditional information to generate a self-supervised speech representation of the data. We utilize a prosody conditional text representation as the prior information. A prosody style representation is extracted from the full-length input speech as a global style embedding. Owing to the semantic information of self-supervised speech representation, we can transfer a prosody style in the TTV framework which is almost irrelevant to the voice style. To increase the linguistic capacity of a semantic representation, a latent representation is fed to the phoneme encoder, and the connectionist temporal classification (CTC) loss is minimized. We found that this could improve text-speech alignment by significantly decreasing the CER and WER of synthetic speech. Furthermore, we use a Transformer-based normalizing flow with AdaLN-Zero for better prosody adaptation.   

\subsection{Speech Super-resolution}

\begin{figure}[t]
    \centering
    {\includegraphics[width=1\columnwidth]{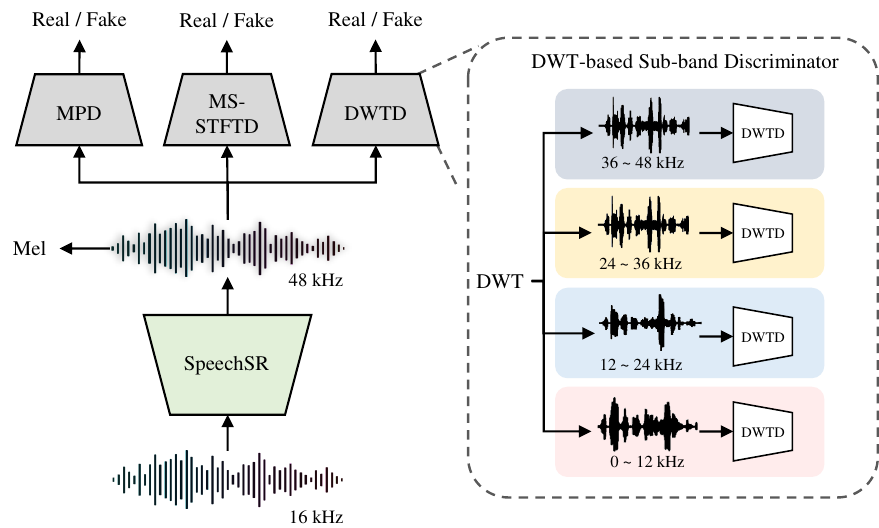}}
    \caption{Speech Super-resolution}
    \label{fg5:sr}
\end{figure}

We use a low-resolution speech dataset to train the model in terms of data availability and efficiency. In this stage, we simply upsample a low-resolution speech waveform to a high-resolution speech waveform from 16 kHz to 48 kHz as illustrated in Fig \ref{fg5:sr}. We use only one anti-aliased multi-periodicity composition (AMP) block of BigVGAN, which consists of a low-pass filter and periodic activation Snake function for inductive bias and anti-aliasing. We further replace a transposed convolution with nearest neighbor (NN) upsampler. Previously, an NN upsampler was shown to alleviate tonal artifacts caused by transposed convolutions. We found that the NN upsampler also reduce the error in the high spectrum compared to the transposed convolution. Based on our hierarchical speech synthesizer, we use MPD and MS-STFTD for high-quality audio synthesis. Additionally, we propose a DWT-based sub-band discriminator (DWTD) to decompose the audio component and reflect the features of each sub-band, respectively as shown in Fig \ref{fg5:sr}. Previously, Fre-GAN \cite{kim21f_interspeech} and Fre-GAN 2 \cite{9746675} already utilize a DWT-based discriminator to replace average polling with a DWT for lossless downsampling. However, in this work, we also decompose a discriminator into sub-band discriminators for each sub-audio ([0 kHz, 12 kHz], [12 kHz, 24 kHz], [24 kHz, 36 kHz], [36 kHz, 48 kHz]), which improves the reconstruction quality for high-frequency bands.        


\subsubsection{Hierarhical Speech Synthesizer}
Dual-audio acoustic encoder consists of waveform audio encoder (wav encoder) and linear spectrogram encoder (spec encoder). The wav encoder consists of AMP blocks of the BigVGAN and downsampling blocks to map the temporal sequence between spectrogram and wav2vec representation. We use downsampling rates of [8,5,4,2] with kernel sizes of [17,10,8,4] and hidden sizes of [16,32,64,128,192]. For the spec encoder, we utilize 16 layers of non-causal WaveNet with hidden size of 192. The HAG consists of source generator and waveform generator. We replace multi-receptive field fusion (MRF) blocks with AMP blocks that has a low-pass filter and periodic activation Snake function for inductive bias and anti-aliasing. For the source generator, we utilize an upsampling rate of [2,2]] with an initial channel of 256. For the waveform generator, we utilize upsampling rates of [4,5,4,2,2] with initial channel of 512. For a discriminator, we utilize a multi-period discriminator (MPD) with the period of [2,3,5,7,11] and a multi-scale STFT-based discriminator (MS-STFTD) with five different sizes of window ([2048,1024,512,256,128]). For the Source-filter encoder, source, filter, and adaptive encoders consist of eight layers of non-causal WaveNet with hidden size of 192. BiT-Flow consists of four residual coupling layers which comprises a previous convolutional networks (preConv), three Transformer blocks, and post convolutional networks (postConv). We adopt convolutional neural networks with a kernel size of five in Transformer blocks for encoding adjacent information and AdaLN-Zero for better voice style adaptation. We utilize a hidden size of 192, a filter size of 768, and two attention heads for the Transformer blocks. We utilize a dropout rate of 0.1 for BiT-Flow. The style encoder consists of two spectral encoders with linear projection and two temporal encoder with 1D convolutional networks, and multi-head self-attention.    \subsection{Model Architecture}
\begin{figure*}[h]
    \centering {\includegraphics[width=0.95\textwidth]{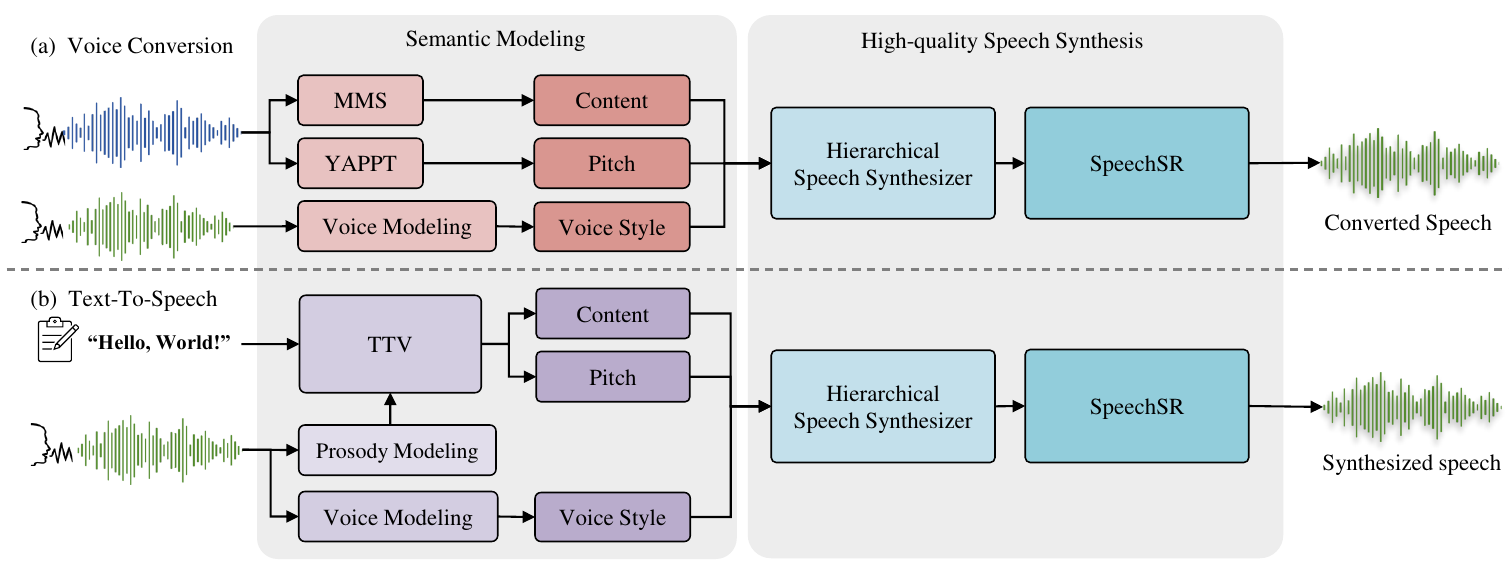}}
    \caption{Inference scenarios for voice conversion and text-to-speech.}
    \label{fg6}
\end{figure*}
\subsubsection{Text-to-Vec}
The content encoder of the TTV consists of 16 layers of non-causal WaveNet with a hidden size of 256 and a kernel size of five. Content decoder consists of eight layers of non-causal WaveNet with hidden size of 512 and kernel size of five. The text encoder is composed of three unconditional Transformer networks and three prosody-conditional Transformer networks with a kernel size of nine, a hidden size of 256 and a filter size of 1024. We utilize a dropout rate of 0.2 for text encoder. T-Flow consists of four residual coupling layers which are composed of a preConv, three Transformer blocks, and postConv. We adopt convolutional neural networks with a kernel size of 5 in Transformer blocks for encoding adjacent information and AdaLN-Zero for better prosody style adaptation. We utilize a hidden size of 256, a filter size of 1024, and four attention heads for T-Flow. We utilize a dropout rate of 0.1 for T-Flow. For the pitch predictor, we utilize the source generator with the same structure as that of HAG.
\subsubsection{SpeechSR}
The SpeechSR consists of a single AMP block with an initial channel of 32 without an upsampling layer. We utilize an NN upsampler for upsampling the hidden representations. For the discriminator, we utilize the MPD with the period of [2,3,5,7,11] and MS-STFTD with six different sizes of window ([4096,2048,1024,512,256,128]). Additionally, we utilize DWTD which has four sub-band discriminators.

\section{Speech Synthesis Tasks}
\subsection{Voice Conversion}
Fig \ref{fg6} illustrates the entire inference pipeline. For voice conversion, we first extract the semantic representation by MMS from the audio at 16 kHz, and F0 using the YAPPT algorithm. Before feeding F0 to Hierarchical Synthesizer, we normalize F0 using the mean and standard deviation of the source speech. Then, we denormalize a normalized F0 by the mean and standard deviation of the target speech. The speech synthesizer synthesizes 16 kHz speech with a target voice style from the target voice prompt. The SpeechSR can upsample the synthesized speech to a high-resolution speech of 48 kHz. For a fair comparison, we do not utilize SpeechSR to evaluate the VC performance.

\subsection{Text-to-Speech}
For text-to-speech, we extract semantic representations from text instead of speech. The TTV can generate a semantic representation with the target prosody from the prosody prompt. The hierarchical speech synthesizer generates speech from semantic representations, and the SpeechSR can upsample it to a high-resolution from 16 kHz to 48 kHz. For a fair comparison, SpeechSR was not used during the TTS evaluation. The prosody and voice styles can be transferred from different target prompts, respectively. 
\begin{figure}[h]
    \centering {\includegraphics[width=1\columnwidth]{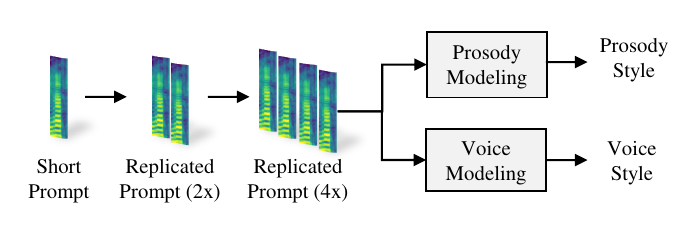}}
    \caption{Style Prompt Replication}
    \label{fg6_spr}
\end{figure}
\subsection{Style Prompt Replication}

We found a simple trick to transfer the style even with a one second speech prompt by introducing style prompt replication (SPR). Similar to the DNA replication, we copy the same sequence of prompt as shown in Fig \ref{fg6_spr}. The replicated prompt by $n$ times is fed to the style encoder to extract the style representation. Specifically, because the prompt style encoder usually encounters a long sequence of prompts over 3s, synthetic speech from short prompts may be generated incorrectly. However, SPR can deceive the style encoder as it seems like long prompts, thus we can synthesize the speech even with 1s speech prompt.   

\section{Experiment and Result}

\begin{table}[h]
\caption{Training Dataset. We utilize public-available speech dataset to train the model. For TTV, we utilize only LibriTTS dataset.}\label{table:dataset}
  \centering 
  \resizebox{0.95\columnwidth}{!}{
  \begin{tabular}{l|c|c|cc}
     \toprule
     Dataset (Subset) & Language & SR (Hz) & Speaker& Hour  \\
    \midrule
    LibriTTS (train-clean-460) & English & 24,000 & 1,151 &  245 \\ 
    LibriTTS (train-other-500)   & English& 24,000  & 1,160 & 310 \\  
    Libri-light* (Medium)   & English & 16,000 & 2,075 & 1,007 \\
    EXPRESSO   & English& 48,000  & 4  & 47 \\
    \midrule
    NIKL  & Korean  & 16,000 & 118  &  98 \\ 
    MSSS & Korean  & 48,000  & 2,791 &  1,101 \\
    \midrule
    Total & - &-& 7,299 & 2,796 \\
    \bottomrule
  \end{tabular}} 
  \vspace{-0.4cm}
\end{table}
\subsection{Dataset}\label{section5-2}
We utilized LibriTTS dataset \cite{zen2019libritts} to train the hierarchical speech synthesizer. First, we trained the model with train-clean subsets of LibriTTS (train-clean-100 and train-clean-360) for a fair comparison. Additionally, we utilized the train-other-500 subsets of LibriTTS for better voice style transfer. Furthermore, we scaled-up the dataset to 1 kh to improve the robustness and diversity, as indicated in TABLE \ref{table:dataset}\footnote{Although we hope to increase the data scale to over 10k Hours, this is the maximum limit in our academic resources.}. For the Libri-light \cite{librilight} and Multi-Speaker Speech Synthesis (MSSS) dataset of AIHub \footnote{\url{https://aihub.or.kr}}, we sampled a small portion of speech from each speaker. We used a EXPRESSO \cite{nguyen23_interspeech} and NIKL\footnote{\url{https://www.nia.or.kr/}}. We downsampled the audio at 16 kHz, and normalized it using a scale of [-0.95, 0.95]. For text-to-vec, we utilized all the train subsets of LibriTTS. For speechSR, we used a VCTK dataset \cite{veaux2017superseded} which has a sampling rate of 48 kHz to compare the models. However, we also trained the model with a large-scale dataset for better speech super-resolution performance by including MSSS dataset, VCTK, and EXPRESSO. 

\subsection{Preprocessing}
We utilized a MMS (0.3B) \cite{pratap2023scaling} which is a pre-trained Wav2Vec 2.0 model with a massively larget-scale cross-lingual speech dataset containing speech dataset of 1000 languages. To map the semantic and acoustic representations, we used a hop size of 320 to extract a linear spectrogram. For the style encoder, we utilized a Mel-spectrogram with 80 bins. We extracted F0 using YAPPT \cite{kasi2002yet} algorithm with a hop of 80. For phoneme transformation, we utilized an International Phonetic Alphabet (IPA) sequence with an open-source Phonemizer \cite{Bernard2021}. Following \cite{lee2022hierspeech}, we did not utilize blank tokens for the target phoneme sequences of the CTC decoder. However, we used blank tokens for the input phoneme sequences.

\subsection{Training}
For reproducibility, we will release the source code and the details of all the hyperparameters will be included at the \url{https://github.com/sh-lee-prml/HierSpeechpp}. We trained HierSpeech++ using the AdamW optimizer \cite{loshchilov2018decoupled} with $\beta_1=0.8$, $\beta_2=0.99$, and weight decay $\lambda=0.01$, and we apply the learning rate schedule by the decay of $0.999^{1/8}$ at an initial learning rate of $1\times10^{-4}$ for the HierSpeech++ with LibriTTS dataset and a batch size of 80 for 1,200k steps on four NVIDIA A6000 GPUs. The final model which is trained with the entire dataset continued to train from HierSpeech++ that was trained with LibriTTS, and the decay decreased by $0.999$. We trained HierSpeech++ with a batch size of 160 for 1,000k steps on eight NVIDIA A6000 GPUs. For the ablation study, the models were trained with a batch size of 80 for 300k steps. We sliced the audio using 61,440 frames for efficient training, and we used windowed generator training for the generator using 9,600 frames. HierSpeech++ consists of 63M parameters for inference and additional 34M parameters only for training. For TTV, we trained the model using the AdamW optimizer \cite{loshchilov2018decoupled} with $\beta_1=0.8$, $\beta_2=0.99$, and weight decay $\lambda=0.01$, and apply the learning rate schedule by the decay of $0.999$ at an initial learning rate of $2\times10^{-4}$ with a batch size of 128 for 950k steps on four NVIDIA A100 GPUs. TTV consists of 107M For SpeechSR, we utilize the same configuration as BigVGAN, and trained the model with a batch size of 128 for 100k steps on four NVIDIA A6000 GPUs.



\begin{table*}[h]

  \caption{Reconstruction results for ablation studies. LT-460 denotes LibriTTS train-clean-100 and 360 subsets, and LT-960 additionally utilize a train-other-500 subset with LT-460. ALL denotes that the model is trained with all dataset in Section \ref{section5-2}. }  \label{table2}
  \centering
  \resizebox{1\textwidth}{!}{
  \begin{tabular}{l|c|cccccc|ccc|cccc}
    \toprule
    Method  & Dataset & AMP & SFE & DAE & T-Flow & Bi-Flow &$\lambda_{bi}$ & Mel & PESQ-wb & PESQ-nb & Pitch & Period.& V/UV F1 & F0$_c$ \\
    \midrule
     &  ALL & \cmark&  \cmark&  \cmark&  \cmark &  \cmark & 0.5 & $\textbf{0.50}$  & $\textbf{2.59}$ & $\textbf{3.20}$ & $\textbf{19.49}$& $\textbf{0.087}$& $\textbf{0.955}$& 0.103\\
  HierSpeech++  &  LT-960 &\cmark&  \cmark&  \cmark&  \cmark &  \cmark & 0.5 & 0.53 & 2.47 & 3.09 & 20.58 & 0.089& 0.952& $\textbf{0.102}$\\
      & LT-460 &\cmark&  \cmark&  \cmark&  \cmark &  \cmark& 0.5 & 0.58 & 2.29 & 2.91 & 24.17 & 0.099& 0.946 & 0.110\\
    \midrule
     & LT-460 &\cmark&  \cmark&  \cmark&  \cmark &  \cmark &0.1 & 0.55 & 2.50 & 3.14 & $\textbf{20.21}$ & 0.090 & 0.950 & $\textbf{0.107}$\\
     & LT-460 &\cmark&  \cmark&  \cmark&  \cmark &  $\cellcolor{gray!25}$ \xmark &  $\cellcolor{gray!25}$ - & $\textbf{0.54}$ & $\textbf{2.55}$ & $\textbf{3.17}$ & 21.23 & $\textbf{0.086}$ & $\textbf{0.953}$ & 0.108 \\
 Ablation Study  & LT-460 &\cmark&  \cmark&  \cmark&  $\cellcolor{gray!25}$ \xmark & $\cellcolor{gray!25}$ \xmark &  $\cellcolor{gray!25}$ - & 0.56 & 2.35 & 3.01 & 23.71 & 0.098 & 0.949 & 0.110\\
    & LT-460 &\cmark&  \cmark& $\cellcolor{gray!25}$ \xmark& $\cellcolor{gray!25}$ \xmark & $\cellcolor{gray!25}$ \xmark & $\cellcolor{gray!25}$ -& 0.63 & 2.29 & 2.85 & 28.68 & 0.102 & 0.947 & 0.133 \\
     & LT-460 &\cmark&  $\cellcolor{gray!25}$ \xmark&  $\cellcolor{gray!25}$ \xmark& $\cellcolor{gray!25}$  \xmark & $\cellcolor{gray!25}$ \xmark &  $\cellcolor{gray!25}$-& 0.63 & 2.23 & 2.82 & 23.85 & 0.106& 0.941 & 0.146 \\
     \midrule
   HierVST   & LT-460 &  $\cellcolor{gray!25}$ \xmark&$\cellcolor{gray!25}$  \xmark& $\cellcolor{gray!25}$ \xmark&$\cellcolor{gray!25}$  \xmark &$\cellcolor{gray!25}$  \xmark & $\cellcolor{gray!25}$ - & 0.66 & 2.18 & 2.77 & 24.53 & 0.104 & 0.944 & 0.125\\
    \bottomrule
  \end{tabular}
  }
\end{table*}

\begin{table*}[h]
 \caption{Resynthesis results for ablation studies}  \label{table3}
  \centering
      \resizebox{1\textwidth}{!}{
  \begin{tabular}{l|c|cccccc|ccc|cccc}
    \toprule
    Method  & Dataset & AMP & SFE & DAE & T-Flow & Bi-Flow &$\lambda_{bi}$ & Mel & PESQ-wb & PESQ-nb & Pitch & Period.& V/UV F1 & F0$_c$\\
    \midrule
    &  ALL &   \cmark&  \cmark&  \cmark&  \cmark & \cmark & 0.5 & $\textbf{0.58}$ & $\textbf{2.32}$ & $\textbf{2.95}$ & $\textbf{20.21}$& $\textbf{0.093}$& $\textbf{0.953}$& $\textbf{0.073}$\\
  HierSpeech++  &    LT-960 &\cmark&  \cmark&  \cmark&  \cmark & \cmark  & 0.5 & 0.61 & 2.23 & 2.84  & 21.29 & 0.090& 0.951 & 0.073\\
      & LT-460 &\cmark&  \cmark&  \cmark&  \cmark &  \cmark  & 0.5 & $\textbf{0.67}$ & 2.04 & 2.63 & 24.26 & 0.107 & 0.943 & $\textbf{0.077}$\\
    \midrule
     & LT-460 &\cmark&  \cmark&  \cmark&  \cmark &  \cmark  & 0.1 &$\textbf{0.67}$ & $\textbf{2.10}$ & $\textbf{2.72}$ & 24.35 & 0.104 & 0.944 & 0.080\\
     & LT-460 &\cmark&  \cmark&  \cmark&  \cmark & $\cellcolor{gray!25}$  \xmark  & $\cellcolor{gray!25}$- & 0.68 & 2.07 & 2.71 & $\textbf{23.56}$ & $\textbf{0.099}$ & $\textbf{0.947}$ & 0.080\\
 Ablation Study     & LT-460 &\cmark&  \cmark&  \cmark& $\cellcolor{gray!25}$ \xmark & $\cellcolor{gray!25}$ \xmark & $\cellcolor{gray!25}$- &  0.68 & 2.05 & 2.68 & 24.85 & 0.106 & 0.943 & 0.080\\
    & LT-460 &\cmark&  \cmark&$\cellcolor{gray!25}$  \xmark& $\cellcolor{gray!25}$ \xmark & $\cellcolor{gray!25}$ \xmark &$\cellcolor{gray!25}$ - &  0.71 & 2.03 & 2.63 & 26.47 & 0.109& 0.944 & 0.111\\
     & LT-460 &\cmark& $\cellcolor{gray!25}$ \xmark& $\cellcolor{gray!25}$ \xmark&  $\cellcolor{gray!25}$\xmark & $\cellcolor{gray!25}$ \xmark & $\cellcolor{gray!25}$- &  0.72 & 1.93 & 2.50 & 34.31 & 0.117& 0.935 & 0.180 \\
     \midrule
   HierVST   & LT-460 & $\cellcolor{gray!25}$\xmark&$\cellcolor{gray!25}$  \xmark& $\cellcolor{gray!25}$ \xmark& $\cellcolor{gray!25}$ \xmark & $\cellcolor{gray!25}$ \xmark& $\cellcolor{gray!25}$- &  0.74 & 1.93 &2.47 & 35.59 & 0.117 & 0.938 & 0.157\\
        
    \bottomrule
  \end{tabular}
  } 
\end{table*}

\begin{table*}[h]
\caption{Voice Conversion results for ablation studies. Params. denotes the number of model parameters.}  \label{table4}
  \centering
      \resizebox{1\textwidth}{!}{
  \begin{tabular}{l|c|cccccc|c|cc|cc|c}
    \toprule
    Method  & Dataset & AMP & SFE & DAE & T-Flow & Bi-Flow &$\lambda_{bi}$ & UTMOS&CER & WER &EER & SECS &  Params. \\
    \midrule
       HierSpeech++  & LT-460 &\cmark&  \cmark&  \cmark&  \cmark &  \cmark & 0.5 & 4.16& 1.09 & 3.17 & $\textbf{4.23}$ & $\textbf{0.856}$ & 63M(+34M) \\
    \midrule
     & LT-460 &\cmark&  \cmark&  \cmark&  \cmark &  \cmark & 0.1 & 4.15&1.15 & 3.49 & 4.82 & 0.849 & 63M(+34M) \\
     & LT-460 &\cmark&  \cmark&  \cmark&  \cmark & $\cellcolor{gray!25}$  \xmark & $\cellcolor{gray!25}$ - & 4.12 &$\textbf{0.83}$ & $\textbf{2.82}$ & 4.86 & 0.847 & 63M(+34M) \\ 
 Ablation Study     & LT-460 &\cmark&  \cmark&  \cmark& $\cellcolor{gray!25}$ \xmark & $\cellcolor{gray!25}$ \xmark & $\cellcolor{gray!25}$ - & 4.13 &  0.95 & 2.98 & 7.50 & 0.845 & 49M(+34M) \\
    & LT-460 &\cmark&  \cmark& $\cellcolor{gray!25}$ \xmark& $\cellcolor{gray!25}$ \xmark & $\cellcolor{gray!25}$ \xmark& $\cellcolor{gray!25}$ - & 4.15& 1.58 & 4.02 & 7.01 & 0.846& 49M(+26M)\\
     & LT-460 &\cmark& $\cellcolor{gray!25}$ \xmark& $\cellcolor{gray!25}$ \xmark& $\cellcolor{gray!25}$ \xmark & $\cellcolor{gray!25}$ \xmark & $\cellcolor{gray!25}$ -&  4.14& 2.01& 4.75& 8.00 & 0.837 & 45M(+21M)\\
     \midrule
   HierVST   & LT-460 & $\cellcolor{gray!25}$\xmark& $\cellcolor{gray!25}$ \xmark& $\cellcolor{gray!25}$ \xmark& $\cellcolor{gray!25}$ \xmark & $\cellcolor{gray!25}$ \xmark& $\cellcolor{gray!25}$ - & 4.08& 2.18 & 5.01 & 9.09 & 0.830 & 45M(+21M) \\ 
    \bottomrule
  \end{tabular}
  }  
\end{table*}

\newpage

\subsection{Evaluation Metrics}
For the reconstruction and resynthesis tasks, we conducted seven objective metrics: a log-scale Mel error distance (Mel), perceptual evaluation of speech quality (PESQ)\footnote{\url{https://github.com/ludlows/PESQ}}, Pitch, periodicity (Period.), voice/unvoice (V/UV) F1 score, and log-scale F0 consistency $F0_c$. We used the official implementation of CARGAN \cite{morrison2022chunked} for pitch, periodicity, and U/UV F1\footnote{\url{https://github.com/descriptinc/cargan}}. For $F0_c$, we calculated the L1 distance between the log-scale ground-truth and the predicted F0 in the HAG. 

For VC, we used two subjective metrics: naturalness mean opinion score (nMOS) and voice similarity MOS (sMOS) with a CI of $95\%$; and three objective metrics for naturalness: UTMOS \cite{saeki22c_interspeech}, character error rate (CER) and word error rate (WER); two objective metrics for similarity: automatic speaker verification equal error rate (EER), and speaker encoder cosine similarity (SECS). We utilized the open-source UTMOS\footnote{\url{https://github.com/tarepan/SpeechMOS}} which is an MOS prediction model for a naturalness metric. Although this can not be considered an absolute evaluation metric, we believe that it is a simple way to estimate the audio quality of synthetic speech. Additionally, this method does not require ground-truth audio or labels to estimate the score. Therefore, we highly recommend using this simple metric during validation by adding a single line. For CER and WER, we utilized the Whisper's official implementation. We used a large model with 1,550 M parameters and calculated the CER and WER after text normalization, as presented in the official implementation. We utilized a pre-trained automatic speaker verification models \cite{kwon2021ins}\footnote{\url{https://github.com/clovaai/voxceleb_trainer}} which was trained with a large-scale speech dataset, VoxCeleb2 \cite{chung2018voxceleb2}. In \cite{chung20b_interspeech}, the effectiveness of metric learning in automatic speaker verification was demonstrated. Furthermore, \cite{kwon2021ins} introduced online data augmentation, which decreased the EER from 2.17$\%$ to 1.17$\%$. In addition, we utilized the pre-trained speaker encoder, Resemblyzer \footnote{\url{https://github.com/resemble-ai/Resemblyzer}} to extract a speaker representation, and we calculated the cosine similarity between the speaker representation of the target speech and synthetic speech.       

For TTS, we additionally utilize a prosody MOS (pMOS). Sixty samples were randomly selected for each model. The nMOS was rated by 10 listeners on a scale of 1-5, and the sMOS and pMOS were rated by 10 listeners on a scale of 1-4. A confidence interval of 95$\%$ was reported for MOS.     
\vspace{-0.2cm}\subsection{Ablation Study}
Before we compare the model with other baselines in the TTS and VC tasks, we conducted ablation studies by comparing Reconstruction\footnote{Reconstruction: Posterior Encoder $\rightarrow$ Generator $\rightarrow$ Audio }, Resynthesis \footnote{Resynthesis: Prior Encoder $\rightarrow$ Generator $\rightarrow$ Audio }, and VC performance to verify the effectiveness of each component in HierSpeech++. First, although previous E2E models have shown high-quality waveform audio generation, the zero-shot speech synthesis performance was considerably low, and some studies must fine-tune or use speaker id for speaker adaptation. Recently, HierVST has significantly improved a voice style transfer performance of the E2E model; therefore so we conduct ablation studies by building up on HierVST.
\\
\textbf{AMP Block  }
We first replaced the MRF block of HiFi-GAN with the AMP of BigVGAN for OOD generation. The AMP improved the performance of all tasks in terms of all metrics without F0 consistency. The results show that the BigVGAN-based HAG performs better but the loss balance may lean toward optimizing the waveform reconstruction rather than F0 prediction; however, the naturalness and similarity of the converted speech improved in terms of all metrics. Specifically, objective naturalness exhibited better UTMOS.\\
\textbf{SF Encoder  }
To address F0 consistency, we utilize an SF encoder (SFE) for a dual-path semantic encoder, which enhances the semantic prior in terms of all metrics. This significantly improved the F0 consistency of inference scenario. It is worth noting that F0 can be manually controlled. \\
\textbf{Dual-audio Encoder  }
We also utilized a dual-audio posterior encoder (DAE) to increase the acoustic capacity of the acoustic representation, which significantly increases the reconstruction performance. Although the linear spectrogram contains useful information for reconstructing a waveform audio, this representation still lacks the ability to reproduce all information; therefore, additional information from waveform audio could complement a wave-level acoustic representation. It is worth noting that the DAE was only utilized during training but significantly improved reconstruction and pronunciation. However, we found that the enhanced acoustic posterior contains a large information resulting in reducing a VC performance. 
\\\textbf{T-Flow  }
To bridge the gap between each representation, we replace a wavenet-based normalizing flow with Transformer-based normalizing flow (T-Flow) using AdaLN-Zero for style adaptation. This also improved the entire performance of all metrics. Moreover, speaker similarity significantly improved. 
\\\textbf{Bi-Flow  }
Moreover, we adopt a bidirectional normalizing flow (Bi-Flow) to reduce the train-inference mismatch problem. The results show that Bi-Flow slightly decreases the reconstruction quality. However, this could regularize a posterior by conditioning the information used in the inference scenario, thereby improving the VC performance. We also found that high weight of Bi-Flow significantly decreased a reconstruction performance, and thus, we use $\lambda$ of 0.5 for weak regularization. 
\\\textbf{Large-scale Data}
In addition, we demonstrated that our model is robust to data scale-up. We did not use any labels to train the model, and only used a low-resolution speech dataset of 16 kHz, which we could obtain simply due to SpeechSR. For scaling-up, we did not conduct any pre-processing to train the model without down-sampling (any to 16 kHz), so there are noisy samples in our dataset however we did not experience any problems.

\begin{table*}[t] 
\caption{Zero-shot VC results on unseen speakers from VCTK dataset. DiffVC$^\spadesuit$ denotes pre-trained model from the official implementation trained with LibriTTS. YourTTS$^\clubsuit$ denotes a pre-trained model from the official implementation trained with LibriTTS, VCTK, and others so the result of YourTTS is not zero-shot VC result.}
\label{table:zero-shotVC_result}\vspace{-0.2cm}
\centering
 \resizebox{1\textwidth}{!}{
  \begin{tabular}{l|c|cc|c|cc|cc}
    \toprule
     Method & Dataset &  nMOS ($\uparrow$) & sMOS ($\uparrow$) &UTMOS ($\uparrow$) & CER ($\downarrow$)  & WER ($\downarrow$) & EER ($\downarrow$) &SECS ($\uparrow$)  \\
      \midrule
         GT   & - &  4.41$\pm$0.06  & 3.89$\pm$0.02  &4.04& 0.21 & 2.17 & - & - \\
      \midrule
      AutoVC \cite{qian2019autovc}& LT-460  & 3.15$\pm$0.09 &2.41$\pm$0.08  & 3.04 &5.14 & 10.55& 37.32 &0.715 \\
      VoiceMixer \cite{lee2021voicemixer}& LT-460 & 3.71$\pm$0.07 & 3.13$\pm$0.06 & 3.19& 1.08 & 3.31 & 20.75&0.797 \\
      \midrule 
      DiffVC$^\spadesuit$ \cite{popov2022diffusionbased}& LT-460  &4.00$\pm$0.07   &3.08$\pm$0.06 &3.49 & 6.86 & 13.77 & 9.25 & 0.826  \\
      Diff-HierVC \cite{choi23d_interspeech}& LT-460  &4.14$\pm$0.06 & 3.50$\pm$0.05  &   3.34& 0.83 & 3.11 & 3.29 & 0.861 \\
      DDDM-VC \cite{choi2023dddm}& LT-460  &   4.13$\pm$0.06 & 3.51$\pm$0.05 & 3.40& 1.77 & 4.35 & 6.49 & 0.858  \\ 
      \midrule
      YourTTS$^\clubsuit$ \cite{casanova2022yourtts} & LT-460, VCTK, Other    & 3.38$\pm$0.08 & 3.13$\pm$0.06 &3.09 & 2.42 & 6.08 & 4.02 & 0.848  \\
      HierVST \cite{lee23i_interspeech}& LT-460 & 4.59$\pm$0.04 &  3.35$\pm$0.06&4.19&1.14&3.46&5.06& 0.850  \\
     HierSpeech++ (Ours) & LT-460& 4.54$\pm$0.05 & 3.63$\pm$0.04 &4.19& 0.90 & 2.99 & 2.50 & 0.862 \\
     HierSpeech++ (Ours) & LT-960& 4.48$\pm$0.05 & 3.62$\pm$0.04 &4.13& 0.79 & 3.01 & 1.27 & 0.875 \\
      HierSpeech++ (Ours) & LT-960, LL, Expresso, MSSS, NIKL &4.44$\pm$0.05 & 3.63$\pm$0.04& 4.13& 0.74 & 2.87 &1.31& 0.883 \\
    \bottomrule
  \end{tabular}} \vspace{-0.2cm}
\end{table*}
\subsection{Zero-shot Voice Conversion}
We compared the voice style transfer performance of HierSpeech++ with other basemodels: 1) AutoVC \cite{qian2019autovc}, which is an autoencoder-based non-autoregressive VC model using an information bottleneck to disentangle the content and style, 2) VoiceMixer \cite{lee2021voicemixer}, which is a GAN-based parallel VC model using similarity-based information bottleneck, 3-5) Diffusion-based models (DiffVC \cite{popov2022diffusionbased}, Diff-HierVC \cite{choi23d_interspeech}, and DDDM-VC \cite{choi2023dddm}), 6) YourTTS \cite{casanova2022yourtts}, VITS-based end-to-end VC models utilizing phoneme sequences to extract content information, 7) HierVST \cite{lee23i_interspeech}, hierspeech-based end-to-end VC model using hierarchical style adaptation. For a fair comparison, we trained all model with the same dataset (LT-460, train-clean-460 subsets of LibriTTS) without YourTTS. We utilized the official implementation of YourTTS which was trained with an additional dataset. We also trained the model with a large-scale dataset such as LT-960, all training subsets of LibriTTS) and additional datasets to verify the effectiveness of scaling-up the dataset. 

For the subjective objective, TABLE \ref{table:zero-shotVC_result} demonstrates that our model significantly improves the naturalness and similarity of the converted speech in terms of nMOS and sMOS. We found that our model with a large-scale dataset showed a better naturalness than ground-truth speech. 

However, the results also showed that increasing the dataset without filtering noisy data slightly decreased an audio quality in terms of nMOS and UTMOS but the similarity increased consistently according to the data scale. In addition, WER also showed better performance than the other models. Furthermore, the results of the similarity measurement show that our models perform better in terms of the EER and SECS. Moreover, the results verified that HierSpeech++ which was trained with a large-scale dataset is a much stronger zero-shot speech synthesizer. 

We will include zero-shot cross-lingual voice style transfer results and additional zero-shot voice conversion results with noisy speech prompts on our demo page. We highly recommend listening to the demo samples and will release the source code of the hierarchical speech synthesizer for a strong zero-shot speech synthesizer. Furthermore, we can also upsample the audio using SpeechSR from 16 kHz to 48 kHz, which can simply improve the perceptual audio quality as described in Section \ref{section:speechsr}. 
\begin{table}[t]
\caption{Temperature parameter search. For TTS, we have two controllable temperature, $T_{ttv}$ of TTV and $T_{h}$ of hierarchical speech synthesizer. We utilize a HierSpeech++ trained with LT-960 for this experiment and fix the random seed. We found that low temperatures improve the robustness of synthetic speech in terms of CER and WER. However, if you hope to generate diverse and expressive speech, you could choose high temperatures. The CER and WER of ground-truth is 2.31 and 4.13, respectively.  UTMOS is presented with standard deviation.}
\label{table:diversity}
\centering
    \resizebox{1\columnwidth}{!}{
    \begin{tabular}{c|c|cccc}
    \toprule
     $T_{ttv}$&$T_{h}$& UTMOS ($\uparrow$)& CER ($\downarrow$)&WER ($\downarrow$) &SECS ($\uparrow$) \\
    \midrule
    0.333&0.333&4.36$\pm$0.25&2.39&4.20&0.907\\
    0.333&0.500&4.35$\pm$0.25\textbf{}&2.35&4.21&0.908\\
    0.333&0.667&4.34$\pm$0.26&2.40&4.31&0.909\\
    0.333&1.000&4.28$\pm$0.29&2.51&4.51&0.911\\
        \midrule
    0.500&0.333&4.36$\pm$0.24&2.17&4.03&0.907\\
    0.500&0.500&4.35$\pm$0.25&2.42&4.35&0.908\\
    0.500&0.667&4.33$\pm$0.26&2.54&4.52&0.909\\
    0.500&1.000&4.27$\pm$0.29&2.64&4.84&0.911\\
        \midrule
    0.667&0.333&4.35$\pm$0.25&2.53&4.45&0.907\\
    0.667&0.500&4.34$\pm$0.25&2.53&4.43&0.908\\
    0.667&0.667&4.32$\pm$0.26&2.56&4.54&0.909\\
    0.667&1.000&4.25$\pm$0.30&3.06&5.33&0.911\\
        \midrule
    1.000&0.333&4.30$\pm$0.27&4.70&6.73&0.905\\
    1.000&0.500&4.29$\pm$0.27&3.65&6.51&0.906\\
    1.000&0.667&4.27$\pm$0.29&4.19&6.71&0.907\\
    1.000&1.000&4.19$\pm$0.32&5.23&7.91&0.909\\
    \bottomrule
  \end{tabular}} \vspace{-0.3cm}
\end{table}
\subsection{High-diversity but High-fidelity Speech Synthesis}
Following Glow-TTS \cite{kim2020glow}, speech with different styles can be synthesized by controlling the temperature parameters in the TTV and hierarchical speech synthesizer. TABLE \ref{table:diversity} shows that lower temperatures ensure the robustness of the synthetic speech in terms of pronunciation. However, the diversity and speaker similarity can be increased by controlling the temperature. Specifically, we found that increasing $T_{ttv}$ improved the similarity of prosody, such as intonation and pronunciation to target prosody prompts and increasing $T_{h}$ improved the similarity of voice style in terms of SECS. However, when the value of $T_{ttv}$ is close to 1, the CER and WER are decreased; therefore, we utilized a value under 1 for robust speech synthesis. In addition, we can synthesize speech differently with different Gaussian noises, and control the duration by multiplying the duration by a specific value.     

\begin{table*}[t]
\caption{Zero-shot TTS results with noisy prompt on unseen speakers from the test-clean subset of LibriTTS. We synthesize all sentences of subset (4,837 samples). For HierSpeech++, we only utilize the text sequences from LibriTTS train-960.}\vspace{-0.1cm}
\label{table:zero-shot-tts1}
\centering
    \resizebox{1\textwidth}{!}{
    \begin{tabular}{l|c|ccc|cccc}
    \toprule
     Method & Dataset &  nMOS ($\uparrow$) &  pMOS ($\uparrow$) & sMOS ($\uparrow$)  & UTMOS ($\uparrow$) & CER ($\downarrow$)  & WER ($\downarrow$) &SECS ($\uparrow$) \\
    \midrule
     GT   & - & 4.32$\pm$0.06 & 3.94$\pm$0.01  & 3.88$\pm$0.02 & 4.04$\pm$0.37& 2.31& 4.13 & -  \\
        \midrule
      YourTTS \cite{casanova2022yourtts}  & LT-460, VCTK, Other & 3.38$\pm$0.07 & 2.69$\pm$0.06& 3.15$\pm$0.06 &3.60$\pm$0.42& 6.16 & 10.53  & 0.831 \\
      HierSpeech \cite{lee2022hierspeech}  & LT-460, VCTK  & 4.02$\pm$0.07& 2.71$\pm$0.07 & 3.35$\pm$0.06&4.05$\pm$0.43& 5.21 & 8.17  & 0.820 \\
      VALL-E-X \cite{wang2023neural} & LT-960, Others (>1,739 hours) & 3.50$\pm$0.08 & 2.75$\pm$0.07 & 3.27$\pm$0.06 & 3.45$\pm$0.62&21.52& 29.33 & 0.865  \\
      XTTS  & Unknown (>49,000 hours) & 3.32$\pm$0.08& 3.57$\pm$0.05 & 3.50$\pm$0.05 & 3.38$\pm$0.44&15.93 & 18.97 & 0.788 \\
      HierSpeech++ (Ours) & LT-460 &4.56$\pm$0.05 & 3.35$\pm$0.06 & 3.70$\pm$0.04 &4.35$\pm$0.25&2.71& 4.59 & 0.899\\
      HierSpeech++ (Ours) & LT-960 &  4.55$\pm$0.04& 3.31$\pm$0.06 & 3.74$\pm$0.04&4.36$\pm$0.25&2.39& 4.20 & 0.907\\
      HierSpeech++ (Ours) & LT-960, LL, Expresso, MSSS, NIKL &4.50$\pm$0.05 & 3.31$\pm$0.06 & 3.72$\pm$0.04 &4.37$\pm$0.23&2.19& 3.87 &  0.911\\
    \bottomrule
  \end{tabular}} \vspace{-0.1cm}
\end{table*}

\begin{table*}[t]
\caption{Zero-shot TTS results with very noisy prompt on unseen speakers from the test-other subset of LibriTTS. We synthesize all sentences of test-other subset (5,120 samples).}\vspace{-0.1cm}
\label{table:zero-shot-tts2}
\centering
    \resizebox{1\textwidth}{!}{
    \begin{tabular}{l|c|ccc|cccc}
    \toprule
     Method & Dataset &  nMOS ($\uparrow$) &  pMOS ($\uparrow$) & sMOS ($\uparrow$)  & UTMOS ($\uparrow$) & CER ($\downarrow$)  & WER ($\downarrow$)  &SECS ($\uparrow$) \\
    \midrule
     GT   & - & 4.15$\pm$0.06  & 3.89$\pm$0.02 & 3.89$\pm$0.02 &3.46$\pm$0.58& 3.37& 7.06& - \\
        \midrule
      YourTTS \cite{casanova2022yourtts}  & LT-460, VCTK, Other & 3.64$\pm$0.08 & 2.73$\pm$0.06 &2.96$\pm$0.06& 3.48$\pm$0.47 & 7.62 & 13.40 & 0.771 \\
      HierSpeech \cite{lee2022hierspeech}  & LT-460, VCTK  & 4.08$\pm$0.07& 2.79$\pm$0.06& 3.26$\pm$0.06& 3.88$\pm$0.45& 6.94 & 7.58  & 0.783 \\ 
      VALL-E-X \cite{wang2023neural} & LT-960, Others (>1,739 hours) & 3.53$\pm$0.08& 2.74$\pm$0.07& 3.14$\pm$0.06 & 3.23$\pm$0.58 & 24.69& 36.42 & 0.839 \\
      XTTS & Unknown (>49,000 hours) & 3.02$\pm$0.09& 3.45$\pm$0.06& 3.48$\pm$0.05 &2.94$\pm$0.56& 35.19 & 38.80  & 0.748 \\
      HierSpeech++ (Ours) & LT-460 & 4.53$\pm$0.05& 3.36$\pm$0.05 & 3.65$\pm$0.04 &4.17$\pm$0.34&3.23& 5.80 & 0.871\\
      HierSpeech++ (Ours) & LT-960 & 4.53$\pm$0.05 &3.38$\pm$0.05& 3.66$\pm$0.04& 4.13$\pm$0.37&2.36& 5.12 & 0.885\\
      HierSpeech++ (Ours) & LT-960, LL, Expresso, MSSS, NIKL & 4.50$\pm$0.05 & 3.35$\pm$0.06 & 3.68$\pm$0.04 & 4.12$\pm$0.38&2.32& 5.15 & 0.887\\
    \bottomrule
  \end{tabular}} \vspace{-0.1cm}
\end{table*}

\subsection{Zero-shot Text-to-Speech}
We compared the zero-shot TTS performance of HierSpeech++ with other baselines: 1) YourTTS, VITS-based end-to-end TTS model, 2) HierSpeech, an end-to-end TTS model using hierarchical VAE, 3) VALL-E-X, a neural codec language models-based multi-lingual zero-shot TTS model and we utilize an unofficial implementation which has an improved audio quality with Vocos decoder, 4) XTTS\footnote{\url{https://github.com/coqui-ai/TTS}}, a TTS product XTTS v1 from Coqui Corp., and XTTS is built on a open-source TTS model, TorToise \cite{betker2023better} which was trained with unprecedented large-scale speech dataset for the first time. For zero-shot TTS, we utilized a noisy speech prompt from the test-clean and test-other subsets of LibriTTS. HierSpeech++ synthesizes the speech with $T_{ttv}$ of 0.333 and $T_{h}$ of 0.333 in TABLE \ref{table:zero-shot-tts1} and \ref{table:zero-shot-tts2}. 

The results demonstrate that our model is a strong zero-shot TTS model in terms of all subjective and objective metrics. We conducted three MOS experiments for naturalness, prosody, and similarity. Our model beats all models significantly, and our model has even surpassed the ground-truth in terms of naturalness. However, XTTS has a better performance in pMOS, and this means learning prosody requires more datasets to improve expressiveness. Although other models show limitations in synthesizing speech with noisy prompts, our model synthesizes a speech robustly. Furthermore, our model has a better CER and WER than ground-truth, and this also demonstrates the robustness of our model. In summary, all results demonstrate the superiority of our model in naturalness, expressiveness, and robustness for zero-shot TTS.

In addition, we could further improve the zero-shot TTS performance by introducing a style prompt replication (SPR) in the following subsection. Note that we do not apply the SPR in TABLE 2-8. The audio could be upsampled to 48 kHz. Lastly, we could also synthesize noise-free speech even with noisy speech. The details will be described in Section \ref{denoise}. 
\newpage
\begin{table}[t]
\caption{Results on the different length of speech prompt. We utilize all sentence over 10s from the test-clean subset of LibriTTS (1,002 samples). SPR denotes style prompt replication and we replicate a short prompt five times for robust style transfer. Because we randomly slice a speech without considering voice/unvoice part, the results of 1s prompts are lower than others. UTMOS is presented with standard deviation.}
\label{table:SPR}
\centering
    \resizebox{1\columnwidth}{!}{
    \begin{tabular}{c|c|cccc}
    \toprule
     Model&Length& UTMOS ($\uparrow$)& CER ($\downarrow$)&WER ($\downarrow$) &SECS ($\uparrow$) \\
     \midrule
     GT &-&4.16$\pm$0.25&1.16&2.32&-\\
        \midrule
    &1&4.07$\pm$0.42&6.23&10.52&0.842 \\
    HierSpeech++ &3&4.38$\pm$0.13&1.76&3.27&0.914\\
    (with SPR)&5&4.41$\pm$0.10&2.00&3.60&0.930\\
    &10&4.44$\pm$0.08&1.34&2.72&0.942\\

    \midrule
    &1& 2.67$\pm$1.05&63.62&86.66&0.757\\
   HierSpeech++ &3&4.30$\pm$0.22&3.87&6.87&0.909\\
   (without SPR) &5&4.40$\pm$0.10&2.59&4.74&0.929\\
    &10&4.44$\pm$0.07&1.78&3.38&0.942\\
    \bottomrule
  \end{tabular}} 
\end{table}
\begin{figure*}[t]
    \centering {\includegraphics[width=0.90\textwidth]{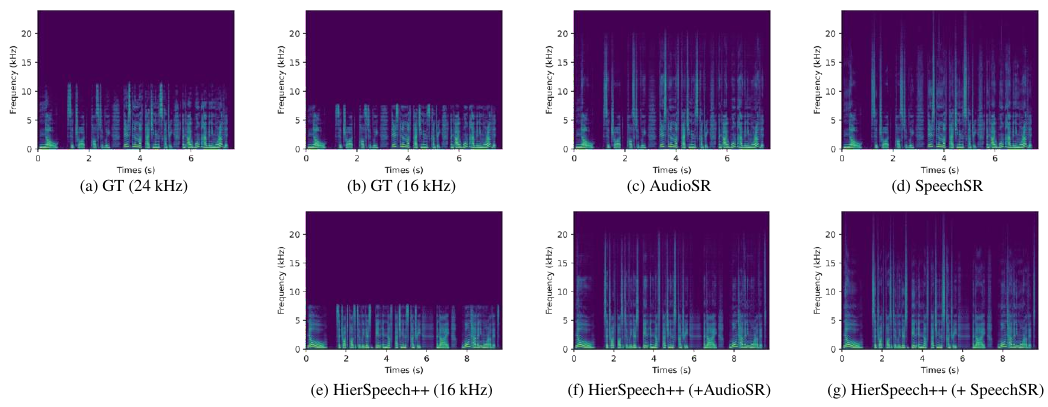}}\vspace{-0.2cm}
    \caption{Spectrograms of GT and speech super-resolution results with AudioSR and SpeechSR (Ours). }\vspace{-0.4cm}
    \label{fg8}
\end{figure*}
\subsection{Zero-shot Text-to-Speech with 1s Prompt}
We compare the performance of zero-shot TTS according to different prompt lengths of 1s, 3s 5s, and 10s. For evaluation, we use all samples over 10s from the test-clean subset of LibriTTS (1,002 samples), and we randomly slice a speech for each prompt length. TABLE \ref{table:SPR} shows that our model has a robust style transfer performance using 3s, 5s, and 10s prompts. However, using 1s prompt could not synthesize a speech well. We can discuss two problems: 1) we do not consider an unvoice part during slicing the speech so some prompts contain only a small portion of speech in their prompt, and we also found that there is no voice part in prompts. 2) we utilize a full-length of prompt during training so synthesizing long sentences may require a long speech prompt for robust speech synthesis, specifically in the prosody encoder. To reduce this problem, we propose a style prompt replication as in section 4.3, and this style prompt replication significantly improves the robustness of TTS. By replicating the prompt like DNA replication, we simply extend a style prompt by $n\times$ and the replicated prompt is fed to the style encoder. This simple trick for style transfer significantly improves the robustness and similarity. With HierSpeech++ using SPR, we could synthesize a speech with only 1s speech prompt even in a zero-shot TTS scenario.   
\begin{table}[t]
\caption{Results of Speech super-resolution on the VCTK dataset.}
\label{table:speechsr}
\centering
    \resizebox{1\columnwidth}{!}{
    \begin{tabular}{c|ccc|cc}
    \toprule
     Model& LSD ($\downarrow$)& LSD-HF ($\downarrow$)& LSD-LF ($\downarrow$)&ViSQOL ($\uparrow$) &PESQ (wb/nb) ($\uparrow$)  \\
     \midrule
     Resampling & 3.41& 4.17& 0.20 & 1.88 &4.64/4.55 \\
      \midrule
    Nu-Wave 2 \cite{han22_interspeech}& 1.01 & 1.18&0.48& 2.62 &4.27/4.44 \\
    UDM+ \cite{10095103}& 1.13 & 1.38& 0.17& 2.73 &4.58/4.55\\
    AudioSR \cite{liu2023audiosr}& 1.36 & 1.61& 0.56& 2.87 &4.39/4.55\\
     \midrule
     SpeechSR (Ours)& 0.82 & 0.98 & 0.30 & 3.34 &4.63/4.55  \\
    w.o DWTD & 0.82 & 0.98& 0.28 & 3.33 & 4.62/4.55 \\
    w.o NNU & 0.83 & 0.99& 0.26 & 3.28 & 4.63/4.55 \\

    \bottomrule
  \end{tabular}} 
\end{table}
\begin{figure}[t]
    \centering {\includegraphics[width=1\columnwidth]{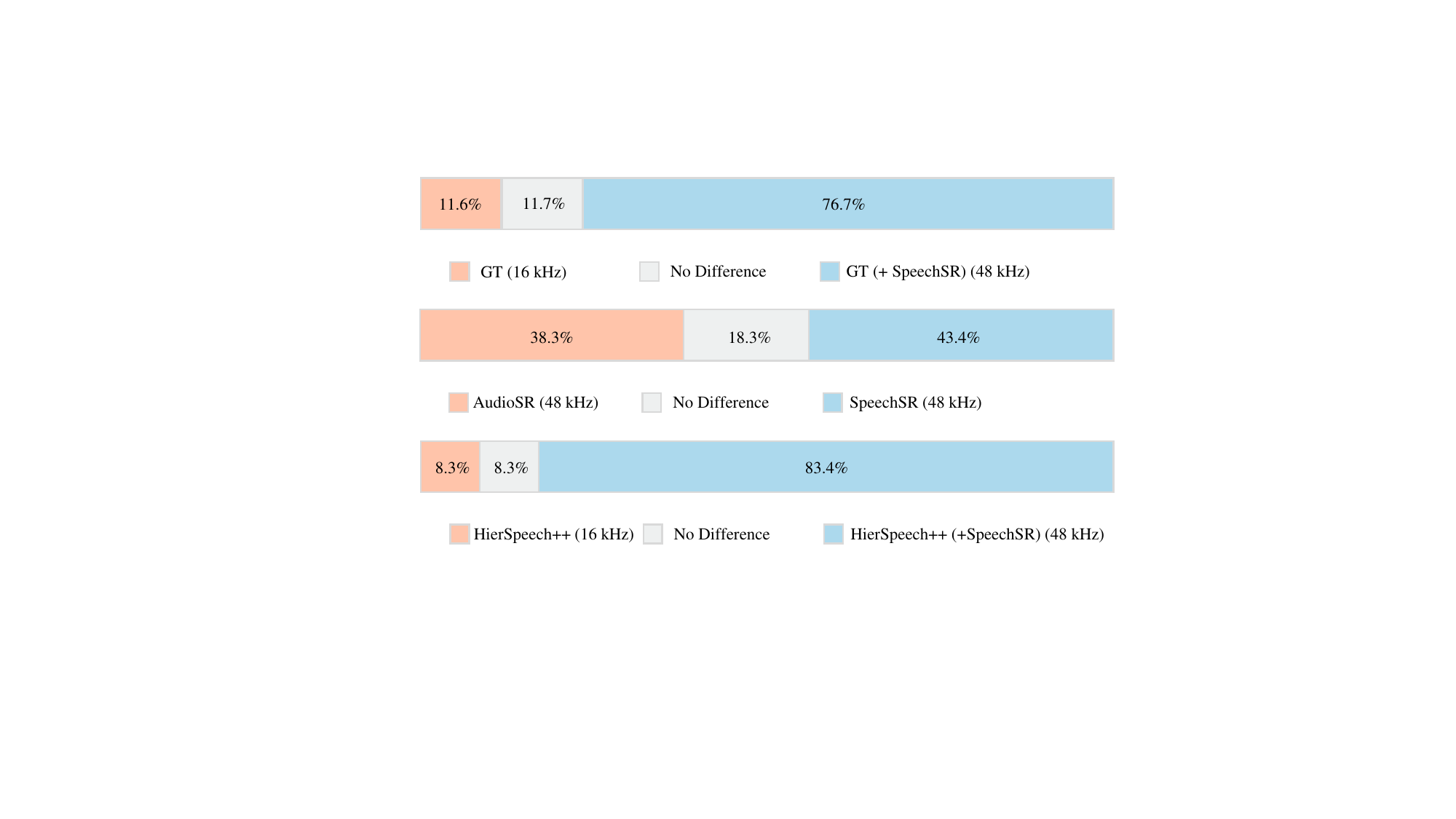}}
    \caption{ABX preference test for SpeechSR.}
    \label{fg9_preference}
\end{figure}

\subsection{Speech Super-resolution}\label{section:speechsr}
We introduced SpeechSR for a simple and efficient speech super-resolution for real-world practical application \cite{9521710}. Because we train a target-specific SpeechSR which can upsample 16 kHz to 48 kHz, our model shows the best performance even with a simple architecture indicated in TABLE \ref{table:speechsr}. For a fair comparison, we trained the model with the VCTK dataset and compare some publicly-available super-resolution models. Although other models perform multi-task super-resolution, we simply focus on 16-48 kHz upsampling for the speech synthesis model. In addition, DTW-based discriminators also improve the super-resolution performance. Furthermore, we scale up the dataset for more robust speech super-resolution and we will release the source code and checkpoint which is trained with the large-scale high-resolution open-source dataset. For the preference test, Fig \ref{fg9_preference} shows that the upsampled speech also shows a better performance than the original speech. Furthermore, it is worth noting that SpeechSR has a 742$\times$ faster inference speed than AudioSR (Speech version) and has a 1,986$\times$ smaller parameter size (SpeechSR has only 0.13M parameters but AudioSR has 258.20M parameters). However, we acknowledge that other models also have a good performance and could upsample any input audio with a sampling rate of 2 kHz to a high-resolution audio with 48 kHz. 

\begin{table}[t]
\caption{Comparison with VALL-E, NaturalSpeech2, and StyleTTS 2. We only compared four samples in the demo page of NaturalSpeech 2 and StyleTTS 2 so this experiment is just for reference.}
\label{table:other_model}
\centering
    \resizebox{1\columnwidth}{!}{
    \begin{tabular}{c|ccc|c}
    \toprule
     Model& UTMOS ($\uparrow$)& CER ($\downarrow$)&WER ($\downarrow$) &SECS (w. Prompt/GT) ($\uparrow$)  \\
     \midrule
     GT &4.13$\pm$0.20&1.58&3.85&0.851/-\\
        \midrule
    Vall-E&3.37$\pm$0.16&2.58&5.10&0.839/0.849 \\
     NaturalSpeech 2&3.79$\pm$0.10&0.91&1.92&0.822/0.837 \\
     StyleTTS 2&4.11$\pm$0.09&2.02&3.46&0.788/0.776 \\
    HierSpeech++&4.20$\pm$0.07&0.0&0.0&0.867/0.810\\
    HierSpeech++ w. SPR&4.26$\pm$0.11&0.0&0.0&0.870/0.822\\
    \bottomrule
  \end{tabular}} 
\end{table}

\subsection{Additional Experiments with Other Baselines}
We compared the zero-shot TTS performance of our model with Vall-E, NaturalSpeech 2, and StyleTTS 2. Because there are no official implementations of them, we utilize the demo samples in NaturalSpeech 2 and StyleTTS 2. We only compared four samples for this experiment. We also added the audio samples to the demo pages. For naturalness, we utilized UTMOS and our model shows a significantly higher score than others. We also compared the similarity with prompt and GT. TABLE \ref{table:other_model} shows that our model has much higher similarity with prompts than others. However, our model has a lower similarity with GT than others. We found that SECS between prompt and GT also shows a low similarity in these samples and this means the prompt and GT have a slightly different style even from the same speaker. In addition, we only utilize four samples. Meanwhile, we thought that the similarity between prompt and generated speech is more important for zero-shot speech synthesis.

\begin{table}[t]
\caption{Results on Speech Prompts with Noise Suppression. HierSpeech++$^\spadesuit$ 
 denotes the cascaded denoising results of Hierspeech++ after speech synthesis. We only utilize the denoised audio as speech prompt for style encoder to extract the denoised style representation.}
\label{table:denoisedstyle}
\centering
    \resizebox{1\columnwidth}{!}{
    \begin{tabular}{c|c|cccc}
    \toprule
     Model& $ratio_d$ & UTMOS ($\uparrow$)& CER ($\downarrow$)&WER ($\downarrow$) &SECS ($\uparrow$) \\
     \midrule
     GT &-&3.46$\pm$0.58&3.37&7.06&-\\
     Denoised GT & - & 3.40$\pm$0.61 &6.91& 10.63  & 0.915 \\
        \midrule
    HierSpeech++  & - & 4.12$\pm$0.38&2.32& 5.15 & 0.887 \\
    HierSpeech++$^\spadesuit$   & - & 4.09$\pm$0.39&3.22& 6.28 & 0.841 \\
      \midrule
   HierSpeech++ &0.1&   4.15$\pm$0.36 &  2.73  &  5.59 & 0.888  \\
   (+Denoised Style)  &0.2&   4.18$\pm$0.35 &   2.63& 5.72  & 0.885  \\
    &0.3&   4.20$\pm$0.33 &  2.42 & 5.18  & 0.881  \\
    &0.4&   4.22$\pm$0.32 &  2.56 & 5.27  & 0.874  \\
    &0.5&   4.23$\pm$0.32 & 2.45  & 5.10  & 0.867  \\
    &0.6&   4.24$\pm$0.31 & 2.56  &  5.28 & 0.859  \\
    &0.7&   4.25$\pm$0.31 & 2.39  &  5.12 & 0.852  \\
    &0.8&   4.25$\pm$0.31 & 2.37& 5.16  & 0.846  \\
    &0.9&   4.25$\pm$0.31 &  3.66 & 6.39  & 0.840  \\
    &1.0&   4.25$\pm$0.32 & 3.99  & 6.88  & 0.834  \\

    \bottomrule
  \end{tabular}} 
\end{table}

\section{Limitation and Quick Fix}\label{denoise}
Although our model improve the zero-shot speech synthesis performance significantly, our model also synthesizes the noisy environmental information from noisy prompt. In this work, we do not disentangle the voice and noise in voice modeling so the model generates a repeated background noise from the global voice style representation. To address this issue, we utilize a denoiser \cite{lu23e_interspeech} to remove a noisy representation in voice style representation. Before fed to style encoder, the audio is fed to denoiser, and the denoised audio is transformed by STFT. Then, the denoised Mel-spectrogram is fed to style encoder. TABLE \ref{table:denoisedstyle} shows that the using denoiser for style encoder simply improve the audio quality in terms of UTMOS. However, the denoised style also decreases the reconstruction quality in terms of CER and WER. As the results of denoised GT degraded all metrics, we found that the denoiser we used also removed the speech part. In this case, the pronunciation of synthetic speech also decreases. To reduce this issue, we interpolate the style representations from the original speech and denoised speech by denoising ratio of $ratio_d$. This simple interpolation significantly improves the audio quality by removing the noisy environmental information without the decrease of CER and WER. For SECS, the denoised speech also shows low SECS and this means that Resemblyzer is also affected by environmental information such as reverberation. It is worth noting that we only utilize a denoiser for style encoder during inference. Meanwhile, using denoiser on the synthetic speech also degrades performance in terms of all metrics as the results of HierSpeech++$^\spadesuit$.

\section{Conclusion}
In this work, we propose HierSpeech++, which achieves a human-level high-quality zero-shot speech synthesis performance. We introduce an efficient and powerful speech synthesis framework by disentangling semantic modeling, speech synthesizer, and speech super-resolution. We thoroughly analyze the components of our model to demonstrate how to achieve a human-level speech synthesis performance even in zero-shot scenarios. Moreover, we simply achieve this performance with a small-scale open-source dataset, LibriTTS. In addition, our model has a significantly faster inference speed than recently proposed zero-shot speech synthesis models. Furthermore, we introduce a style prompt replication for 1s voice cloning, and noise-free speech synthesis by adopting a denoised style prompt. Furthermore, SpeechSR simply upsamples the audio to 48 kHz for high-resolution audio generation. We will release the source code and checkpoint of all components including TTV, hierarchical speech synthesizer, and SpeechSR. For future works, we will extend the model to cross-lingual and emotion-controllable speech synthesis models by utilizing the pre-trained models such as \cite{10089511}. Furthermore, we see that our hierarchical speech synthesis framework could be adopted to a speech-to-speech translation system by introducing non-autoregressive generation \cite{10129160}. 

\ifCLASSOPTIONcompsoc
  \section*{Acknowledgments}
\else
  \section*{Acknowledgment}
\fi
We'd like to thank Hongsun Yang for helpful discussions and contributions to our work. This study used an open-source korean speech datasets, NIKL dataset from the NIA and multi-speaker speech synthesis (MSSS) dataset from the AIHub.

\ifCLASSOPTIONcaptionsoff
  \newpage
\fi



%
\bibliographystyle{abbrv}
\bibliography{biblio}

\begin{thebibliography}{10}

\bibitem{babu22_interspeech}
A.~Babu, C.~Wang, A.~Tjandra, K.~Lakhotia, Q.~Xu, N.~Goyal, K.~Singh, P.~{von
  Platen}, Y.~Saraf, J.~Pino, A.~Baevski, A.~Conneau, and M.~Auli.
\newblock {XLS-R}: Self-supervised cross-lingual speech representation learning
  at scale.
\newblock In {\em Proc. Interspeech}, pages 2278--2282, 2022.

\bibitem{baevski2020wav2vec}
A.~Baevski, Y.~Zhou, A.~Mohamed, and M.~Auli.
\newblock wav2vec 2.0: A framework for self-supervised learning of speech
  representations.
\newblock {\em Proc. Adv. Neural Inf. Process. Syst.}, 33:12449--12460, 2020.

\bibitem{6472238}
Y.~Bengio, A.~Courville, and P.~Vincent.
\newblock Representation learning: A review and new perspectives.
\newblock {\em IEEE Trans. Pattern Anal. Mach. Intell.}, 35(8):1798--1828,
  2013.

\bibitem{Bernard2021}
M.~Bernard and H.~Titeux.
\newblock Phonemizer: Text to phones transcription for multiple languages in
  python.
\newblock {\em Journal of Open Source Software}, 6(68):3958, 2021.

\bibitem{betker2023better}
J.~Betker.
\newblock Better speech synthesis through scaling.
\newblock {\em arXiv preprint arXiv:2305.07243}, 2023.

\bibitem{borsos2023soundstorm}
Z.~Borsos, M.~Sharifi, D.~Vincent, E.~Kharitonov, N.~Zeghidour, and
  M.~Tagliasacchi.
\newblock Soundstorm: Efficient parallel audio generation.
\newblock {\em arXiv preprint arXiv:2305.09636}, 2023.

\bibitem{casanova2022yourtts}
E.~Casanova, J.~Weber, C.~D. Shulby, A.~C. Junior, E.~G{\"o}lge, and M.~A.
  Ponti.
\newblock {YourTTS: Towards Zero-Shot Multi-Speaker TTS and Zero-Shot Voice
  Conversion for everyone}.
\newblock In {\em Proc. Int. Conf. on Mach. Learn.}, pages 2709--2720. PMLR,
  2022.

\bibitem{chen2021adaspeech}
M.~Chen, X.~Tan, B.~Li, Y.~Liu, T.~Qin, S.~Zhao, and T.-Y. Liu.
\newblock Adaspeech: Adaptive text to speech for custom voice.
\newblock {\em arXiv preprint arXiv:2103.00993}, 2021.

\bibitem{choi2022nansy++}
H.-S. Choi, J.~Yang, J.~Lee, and H.~Kim.
\newblock Nansy++: Unified voice synthesis with neural analysis and synthesis.
\newblock {\em arXiv preprint arXiv:2211.09407}, 2022.

\bibitem{choi2023dddm}
H.-Y. Choi, S.-H. Lee, and S.-W. Lee.
\newblock Dddm-vc: Decoupled denoising diffusion models with disentangled
  representation and prior mixup for verified robust voice conversion.
\newblock {\em arXiv preprint arXiv:2305.15816}, 2023.

\bibitem{choi23d_interspeech}
H.-Y. Choi, S.-H. Lee, and S.-W. Lee.
\newblock {Diff-HierVC: Diffusion-based Hierarchical Voice Conversion with
  Robust Pitch Generation and Masked Prior for Zero-shot Speaker Adaptation}.
\newblock In {\em Proc. INTERSPEECH}, pages 2283--2287, 2023.

\bibitem{chung21_interspeech}
H.~Chung, S.-H. Lee, and S.-W. Lee.
\newblock {Reinforce-Aligner: Reinforcement Alignment Search for Robust
  End-to-End Text-to-Speech}.
\newblock In {\em Proc. Interspeech}, pages 3635--3639, 2021.

\bibitem{chung20b_interspeech}
J.~S. Chung, J.~Huh, S.~Mun, M.~Lee, H.-S. Heo, S.~Choe, C.~Ham, S.~Jung, B.-J.
  Lee, and I.~Han.
\newblock {In Defence of Metric Learning for Speaker Recognition}.
\newblock In {\em Proc. Interspeech 2020}, pages 2977--2981, 2020.

\bibitem{chung2018voxceleb2}
J.~S. Chung, A.~Nagrani, and A.~Zisserman.
\newblock Voxceleb2: Deep speaker recognition.
\newblock In {\em Proc. Interspeech}, pages 1086--1090, 2018.

\bibitem{defossez2022high}
A.~D{\'e}fossez, J.~Copet, G.~Synnaeve, and Y.~Adi.
\newblock High fidelity neural audio compression.
\newblock {\em arXiv preprint arXiv:2210.13438}, 2022.

\bibitem{dfossez2023high}
A.~D{\'e}fossez, J.~Copet, G.~Synnaeve, and Y.~Adi.
\newblock High fidelity neural audio compression.
\newblock {\em Trans. Mach. Learn. Research}, 2023.

\bibitem{du2022vqtts}
C.~Du, Y.~Guo, X.~Chen, and K.~Yu.
\newblock Vqtts: High-fidelity text-to-speech synthesis with self-supervised vq
  acoustic feature.
\newblock {\em arXiv preprint arXiv:2204.00768}, 2022.

\bibitem{10229489}
C.~Du, Y.~Guo, X.~Chen, and K.~Yu.
\newblock Speaker adaptive text-to-speech with timbre-normalized
  vector-quantized feature.
\newblock {\em IEEE/ACM Trans. Audio, Speech, Lang. Process.}, pages 1--12,
  2023.

\bibitem{guo2023prompttts}
Z.~Guo, Y.~Leng, Y.~Wu, S.~Zhao, and X.~Tan.
\newblock Prompttts: Controllable text-to-speech with text descriptions.
\newblock In {\em IEEE Int. Conf. Acoust., Speech, Signal Process.}, pages
  1--5. IEEE, 2023.

\bibitem{9716741}
K.~Han, Y.~Wang, H.~Chen, X.~Chen, J.~Guo, Z.~Liu, Y.~Tang, A.~Xiao, C.~Xu,
  Y.~Xu, Z.~Yang, Y.~Zhang, and D.~Tao.
\newblock A survey on vision transformer.
\newblock {\em IEEE Trans. Pattern Anal. Mach. Intell.}, 45(1):87--110, 2023.

\bibitem{han22_interspeech}
S.~Han and J.~Lee.
\newblock {NU-Wave 2: A General Neural Audio Upsampling Model for Various
  Sampling Rates}.
\newblock In {\em Proc. Interspeech}, pages 4401--4405, 2022.

\bibitem{huang2022generspeech}
R.~Huang, Y.~Ren, J.~Liu, C.~Cui, and Z.~Zhao.
\newblock Generspeech: Towards style transfer for generalizable out-of-domain
  text-to-speech.
\newblock In {\em Proc. Adv. Neural Inf. Process. Syst.}, 2022.

\bibitem{huang2023make}
R.~Huang, C.~Zhang, Y.~Wang, D.~Yang, L.~Liu, Z.~Ye, Z.~Jiang, C.~Weng,
  Z.~Zhao, and D.~Yu.
\newblock Make-a-voice: Unified voice synthesis with discrete representation.
\newblock {\em arXiv preprint arXiv:2305.19269}, 2023.

\bibitem{hwang2023hiddensinger}
J.-S. Hwang, S.-H. Lee, and S.-W. Lee.
\newblock Hiddensinger: High-quality singing voice synthesis via neural audio
  codec and latent diffusion models.
\newblock {\em arXiv preprint arXiv:2306.06814}, 2023.

\bibitem{jia2018transfer}
Y.~Jia, Y.~Zhang, R.~Weiss, Q.~Wang, J.~Shen, F.~Ren, P.~Nguyen, R.~Pang,
  I.~Lopez~Moreno, Y.~Wu, et~al.
\newblock Transfer learning from speaker verification to multispeaker
  text-to-speech synthesis.
\newblock {\em Proc. Adv. Neural Inf. Process. Syst.}, 31, 2018.

\bibitem{jiang2023mega2}
Z.~Jiang, J.~Liu, Y.~Ren, J.~He, C.~Zhang, Z.~Ye, P.~Wei, C.~Wang, X.~Yin,
  Z.~Ma, et~al.
\newblock Mega-tts 2: Zero-shot text-to-speech with arbitrary length speech
  prompts.
\newblock {\em arXiv preprint arXiv:2307.07218}, 2023.

\bibitem{librilight}
J.~{Kahn}, M.~{Rivière}, W.~{Zheng}, E.~{Kharitonov}, Q.~{Xu}, P.~E.
  {Mazaré}, J.~{Karadayi}, V.~{Liptchinsky}, R.~{Collobert}, C.~{Fuegen},
  T.~{Likhomanenko}, G.~{Synnaeve}, A.~{Joulin}, A.~{Mohamed}, and E.~{Dupoux}.
\newblock Libri-light: A benchmark for asr with limited or no supervision.
\newblock In {\em ICASSP 2020 - 2020 IEEE Int. Conf. on Acoustics, Speech and
  Signal Processing (ICASSP)}, pages 7669--7673, 2020.

\bibitem{9306875}
H.~Kameoka, W.-C. Huang, K.~Tanaka, T.~Kaneko, N.~Hojo, and T.~Toda.
\newblock Many-to-many voice transformer network.
\newblock {\em IEEE/ACM Trans. Audio, Speech, Lang. Process.}, 29:656--670,
  2021.

\bibitem{kang2023grad}
M.~Kang, D.~Min, and S.~J. Hwang.
\newblock Grad-stylespeech: Any-speaker adaptive text-to-speech synthesis with
  diffusion models.
\newblock In {\em IEEE Int. Conf. Acoust., Speech, Signal Process.}, pages
  1--5. IEEE, 2023.

\bibitem{kasi2002yet}
K.~Kasi and S.~A. Zahorian.
\newblock Yet another algorithm for pitch tracking.
\newblock In {\em IEEE Int. Conf. Acoust., Speech, Signal Process.}, volume~1,
  pages I--361, 2002.

\bibitem{kharitonov2023speak}
E.~Kharitonov, D.~Vincent, Z.~Borsos, R.~Marinier, S.~Girgin, O.~Pietquin,
  M.~Sharifi, M.~Tagliasacchi, and N.~Zeghidour.
\newblock Speak, read and prompt: High-fidelity text-to-speech with minimal
  supervision.
\newblock {\em arXiv preprint arXiv:2302.03540}, 2023.

\bibitem{kim2023unitspeech}
H.~Kim, S.~Kim, J.~Yeom, and S.~Yoon.
\newblock Unitspeech: Speaker-adaptive speech synthesis with untranscribed
  data.
\newblock {\em arXiv preprint arXiv:2306.16083}, 2023.

\bibitem{kim2022guided}
H.~Kim, S.~Kim, and S.~Yoon.
\newblock Guided-{TTS}: A diffusion model for text-to-speech via classifier
  guidance.
\newblock In {\em Proc. Int. Conf. on Mach. Learn.}, pages 11119--11133, 2022.

\bibitem{kim2020glow}
J.~Kim, S.~Kim, J.~Kong, and S.~Yoon.
\newblock Glow-{TTS}: A generative flow for text-to-speech via monotonic
  alignment search.
\newblock {\em Proc. Adv. Neural Inf. Process. Syst.}, 33:8067--8077, 2020.

\bibitem{kim2021conditional}
J.~Kim, J.~Kong, and J.~Son.
\newblock Conditional variational autoencoder with adversarial learning for
  end-to-end text-to-speech.
\newblock In {\em Proc. Int. Conf. on Mach. Learn.}, pages 5530--5540. PMLR,
  2021.

\bibitem{kim21f_interspeech}
J.-H. Kim, S.-H. Lee, J.-H. Lee, and S.-W. Lee.
\newblock {Fre-GAN: Adversarial Frequency-Consistent Audio Synthesis}.
\newblock In {\em Proc. Interspeech}, pages 2197--2201, 2021.

\bibitem{kim22c_interspeech}
M.~Kim, M.~Jeong, B.~J. Choi, S.~Ahn, J.~Y. Lee, and N.~S. Kim.
\newblock {Transfer Learning Framework for Low-Resource Text-to-Speech using a
  Large-Scale Unlabeled Speech Corpus}.
\newblock In {\em Proc. Interspeech}, pages 788--792, 2022.

\bibitem{kim2022guided2}
S.~Kim, H.~Kim, and S.~Yoon.
\newblock Guided-{TTS} 2: A diffusion model for high-quality adaptive
  text-to-speech with untranscribed data.
\newblock {\em arXiv preprint arXiv:2205.15370}, 2022.

\bibitem{kim2023pflow}
S.~Kim, K.~J. Shih, R.~Badlani, J.~F. Santos, E.~Bakhturina, M.~T. Desta,
  R.~Valle, S.~Yoon, and B.~Catanzaro.
\newblock P-flow: A fast and data-efficient zero-shot {TTS} through speech
  prompting.
\newblock In {\em Proc. Adv. Neural Inf. Process. Syst.}, 2023.

\bibitem{kong2020hifi}
J.~Kong, J.~Kim, and J.~Bae.
\newblock Hi{F}i-{GAN}: Generative adversarial networks for efficient and high
  fidelity speech synthesis.
\newblock {\em Proc. Adv. Neural Inf. Process. Syst.}, 33:17022--17033, 2020.

\bibitem{kumar2023highfidelity}
R.~Kumar, P.~Seetharaman, A.~Luebs, I.~Kumar, and K.~Kumar.
\newblock High-fidelity audio compression with improved {RVQGAN}.
\newblock In {\em Proc. Adv. Neural Inf. Process. Syst.}, 2023.

\bibitem{kwon2021ins}
Y.~Kwon, H.~S. Heo, B.-J. Lee, and J.~S. Chung.
\newblock The ins and outs of speaker recognition: lessons from {VoxSRC} 2020.
\newblock In {\em IEEE Int. Conf. Acoust., Speech, Signal Process.}, 2021.

\bibitem{le2023voicebox}
M.~Le, A.~Vyas, B.~Shi, B.~Karrer, L.~Sari, R.~Moritz, M.~Williamson,
  V.~Manohar, Y.~Adi, J.~Mahadeokar, and W.-N. Hsu.
\newblock Voicebox: Text-guided multilingual universal speech generation at
  scale.
\newblock In {\em Proc. Adv. Neural Inf. Process. Syst.}, 2023.

\bibitem{lee2022pvae}
J.-H. Lee, S.-H. Lee, J.-H. Kim, and S.-W. Lee.
\newblock Pvae-tts: Adaptive text-to-speech via progressive style adaptation.
\newblock In {\em IEEE Int. Conf. Acoust., Speech, Signal Process.}, pages
  6312--6316. IEEE, 2022.

\bibitem{lee23i_interspeech}
S.-H. Lee, H.-Y. Choi, H.-S. Oh, and S.-W. Lee.
\newblock {HierVST: Hierarchical Adaptive Zero-shot Voice Style Transfer}.
\newblock In {\em Proc. Interspeech}, pages 4439--4443, 2023.

\bibitem{lee2021voicemixer}
S.-H. Lee, J.-H. Kim, H.~Chung, and S.-W. Lee.
\newblock Voice{M}ixer: Adversarial voice style mixup.
\newblock {\em Proc. Adv. Neural Inf. Process. Syst.}, 34:294--308, 2021.

\bibitem{9746675}
S.-H. Lee, J.-H. Kim, K.-E. Lee, and S.-W. Lee.
\newblock Fre-gan 2: Fast and efficient frequency-consistent audio synthesis.
\newblock In {\em IEEE Int. Conf. Acoust., Speech, Signal Process.}, pages
  6192--6196, 2022.

\bibitem{lee2022hierspeech}
S.-H. Lee, S.-B. Kim, J.-H. Lee, E.~Song, M.-J. Hwang, and S.-W. Lee.
\newblock Hier{S}peech: Bridging the gap between text and speech by
  hierarchical variational inference using self-supervised representations for
  speech synthesis.
\newblock In {\em Proc. Adv. Neural Inf. Process. Syst.}, 2022.

\bibitem{9729483}
S.-H. Lee, H.-R. Noh, W.-J. Nam, and S.-W. Lee.
\newblock Duration controllable voice conversion via phoneme-based information
  bottleneck.
\newblock {\em IEEE/ACM Trans. Audio, Speech, Lang. Process.}, 30:1173--1183,
  2022.

\bibitem{lee2021multi}
S.-H. Lee, H.-W. Yoon, H.-R. Noh, J.-H. Kim, and S.-W. Lee.
\newblock Multi-spectrogan: High-diversity and high-fidelity spectrogram
  generation with adversarial style combination for speech synthesis.
\newblock In {\em Proc. AAAI Conf. Artif. Intell.}, volume~35, pages
  13198--13206, 2021.

\bibitem{lee2019robust}
Y.~Lee and T.~Kim.
\newblock Robust and fine-grained prosody control of end-to-end speech
  synthesis.
\newblock In {\em IEEE Int. Conf. Acoust., Speech, Signal Process.}, pages
  5911--5915. IEEE, 2019.

\bibitem{leng2023prompttts}
Y.~Leng, Z.~Guo, K.~Shen, X.~Tan, Z.~Ju, Y.~Liu, Y.~Liu, D.~Yang, L.~Zhang,
  K.~Song, et~al.
\newblock Prompttts 2: Describing and generating voices with text prompt.
\newblock {\em arXiv preprint arXiv:2309.02285}, 2023.

\bibitem{li2019neural}
N.~Li, S.~Liu, Y.~Liu, S.~Zhao, and M.~Liu.
\newblock Neural speech synthesis with transformer network.
\newblock In {\em Proc. AAAI Conf. Artif. Intell.}, volume~33, pages
  6706--6713, 2019.

\bibitem{li2023styletts}
Y.~A. Li, C.~Han, V.~S. Raghavan, G.~Mischler, and N.~Mesgarani.
\newblock Styletts 2: Towards human-level text-to-speech through style
  diffusion and adversarial training with large speech language models.
\newblock {\em arXiv preprint arXiv:2306.07691}, 2023.

\bibitem{liu2023audiosr}
H.~Liu, K.~Chen, Q.~Tian, W.~Wang, and M.~D. Plumbley.
\newblock Audiosr: Versatile audio super-resolution at scale.
\newblock {\em arXiv preprint arXiv:2309.07314}, 2023.

\bibitem{9420297}
S.~Liu, Y.~Cao, D.~Wang, X.~Wu, X.~Liu, and H.~Meng.
\newblock Any-to-many voice conversion with location-relative
  sequence-to-sequence modeling.
\newblock {\em IEEE/ACM Trans. Audio, Speech, Lang. Process.}, 29:1717--1728,
  2021.

\bibitem{loshchilov2018decoupled}
I.~Loshchilov and F.~Hutter.
\newblock Decoupled weight decay regularization.
\newblock In {\em Proc. Int. Conf. Learn. Representations}, 2019.

\bibitem{lu23e_interspeech}
Y.-X. Lu, Y.~Ai, and Z.-H. Ling.
\newblock {MP-SENet: A Speech Enhancement Model with Parallel Denoising of
  Magnitude and Phase Spectra}.
\newblock In {\em Proc. INTERSPEECH}, pages 3834--3838, 2023.

\bibitem{min2021meta}
D.~Min, D.~B. Lee, E.~Yang, and S.~J. Hwang.
\newblock Meta-stylespeech: Multi-speaker adaptive text-to-speech generation.
\newblock In {\em Proc. Int. Conf. on Mach. Learn.}, pages 7748--7759. PMLR,
  2021.

\bibitem{morrison2022chunked}
M.~Morrison, R.~Kumar, K.~Kumar, P.~Seetharaman, A.~Courville, and Y.~Bengio.
\newblock Chunked autoregressive {GAN} for conditional waveform synthesis.
\newblock In {\em Proc. Int. Conf. Learn. Representations}, 2022.

\bibitem{nguyen23_interspeech}
T.~A. Nguyen, W.-N. Hsu, A.~D'Avirro, B.~Shi, I.~Gat, M.~Fazel-Zarani,
  T.~Remez, J.~Copet, G.~Synnaeve, M.~Hassid, F.~Kreuk, Y.~Adi, and E.~Dupoux.
\newblock {Expresso: A Benchmark and Analysis of Discrete Expressive Speech
  Resynthesis}.
\newblock In {\em Proc. INTERSPEECH}, pages 4823--4827, 2023.

\bibitem{peebles2023scalable}
W.~Peebles and S.~Xie.
\newblock Scalable diffusion models with transformers.
\newblock In {\em Proc. of the IEEE/CVF Int. Conf. on Computer Vision}, pages
  4195--4205, 2023.

\bibitem{popov2021grad}
V.~Popov, I.~Vovk, V.~Gogoryan, T.~Sadekova, and M.~Kudinov.
\newblock Grad-{TTS}: A diffusion probabilistic model for text-to-speech.
\newblock In {\em Proc. Int. Conf. on Mach. Learn.}, pages 8599--8608, 2021.

\bibitem{popov2022diffusionbased}
V.~Popov, I.~Vovk, V.~Gogoryan, T.~Sadekova, M.~S. Kudinov, and J.~Wei.
\newblock Diffusion-based voice conversion with fast maximum likelihood
  sampling scheme.
\newblock In {\em Proc. Int. Conf. Learn. Representations}, 2022.

\bibitem{pratap2023scaling}
V.~Pratap, A.~Tjandra, B.~Shi, P.~Tomasello, A.~Babu, S.~Kundu, A.~Elkahky,
  Z.~Ni, A.~Vyas, M.~Fazel-Zarandi, et~al.
\newblock Scaling speech technology to 1,000+ languages.
\newblock {\em arXiv preprint arXiv:2305.13516}, 2023.

\bibitem{qian2019autovc}
K.~Qian, Y.~Zhang, S.~Chang, X.~Yang, and M.~Hasegawa-Johnson.
\newblock Auto{VC}: Zero-shot voice style transfer with only autoencoder loss.
\newblock In {\em Proc. Int. Conf. on Mach. Learn.}, pages 5210--5219, 2019.

\bibitem{ren2020fastspeech}
Y.~Ren, C.~Hu, X.~Tan, T.~Qin, S.~Zhao, Z.~Zhao, and T.-Y. Liu.
\newblock Fastspeech 2: Fast and high-quality end-to-end text to speech.
\newblock {\em arXiv preprint arXiv:2006.04558}, 2020.

\bibitem{ren2019fastspeech}
Y.~Ren, Y.~Ruan, X.~Tan, T.~Qin, S.~Zhao, Z.~Zhao, and T.-Y. Liu.
\newblock Fastspeech: Fast, robust and controllable text to speech.
\newblock {\em Proc. Adv. Neural Inf. Process. Syst.}, 32, 2019.

\bibitem{saeki22c_interspeech}
T.~Saeki, D.~Xin, W.~Nakata, T.~Koriyama, S.~Takamichi, and H.~Saruwatari.
\newblock {UTMOS: UTokyo-SaruLab System for VoiceMOS Challenge 2022}.
\newblock In {\em Proc. Interspeech 2022}, pages 4521--4525, 2022.

\bibitem{shen2023naturalspeech}
K.~Shen, Z.~Ju, X.~Tan, Y.~Liu, Y.~Leng, L.~He, T.~Qin, S.~Zhao, and J.~Bian.
\newblock Naturalspeech 2: Latent diffusion models are natural and zero-shot
  speech and singing synthesizers.
\newblock {\em arXiv preprint arXiv:2304.09116}, 2023.

\bibitem{siuzdak2022wavthruvec}
H.~Siuzdak, P.~Dura, P.~van Rijn, and N.~Jacoby.
\newblock Wavthruvec: Latent speech representation as intermediate features for
  neural speech synthesis.
\newblock {\em arXiv preprint arXiv:2203.16930}, 2022.

\bibitem{skerry2018towards}
R.~Skerry-Ryan, E.~Battenberg, Y.~Xiao, Y.~Wang, D.~Stanton, J.~Shor, R.~Weiss,
  R.~Clark, and R.~A. Saurous.
\newblock Towards end-to-end prosody transfer for expressive speech synthesis
  with tacotron.
\newblock In {\em Proc. Int. Conf. on Mach. Learn.}, pages 4693--4702. PMLR,
  2018.

\bibitem{9521710}
S.~Son, J.~Kim, W.-S. Lai, M.-H. Yang, and K.~M. Lee.
\newblock Toward real-world super-resolution via adaptive downsampling models.
\newblock {\em IEEE Trans. Pattern Anal. Mach. Intell.}, 44(11):8657--8670,
  2022.

\bibitem{10073591}
H.~Sun, D.~Wang, L.~Li, C.~Chen, and T.~F. Zheng.
\newblock Random cycle loss and its application to voice conversion.
\newblock {\em IEEE Trans. Pattern Anal. Mach. Intell.}, 45(8):10331--10345,
  2023.

\bibitem{tan2022naturalspeech}
X.~Tan, J.~Chen, H.~Liu, J.~Cong, C.~Zhang, Y.~Liu, X.~Wang, Y.~Leng, Y.~Yi,
  L.~He, et~al.
\newblock Naturalspeech: End-to-end text to speech synthesis with human-level
  quality.
\newblock {\em arXiv preprint arXiv:2205.04421}, 2022.

\bibitem{veaux2017superseded}
C.~Veaux, J.~Yamagishi, K.~MacDonald, et~al.
\newblock {Superseded-CSTR VCTK corpus: English multi-speaker corpus for CSTR
  voice cloning toolkit}.
\newblock 2017.

\bibitem{10089511}
J.~Wagner, A.~Triantafyllopoulos, H.~Wierstorf, M.~Schmitt, F.~Burkhardt,
  F.~Eyben, and B.~W. Schuller.
\newblock Dawn of the transformer era in speech emotion recognition: Closing
  the valence gap.
\newblock {\em IEEE Trans. Pattern Anal. Mach. Intell.}, 45(9):10745--10759,
  2023.

\bibitem{wang2023neural}
C.~Wang, S.~Chen, Y.~Wu, Z.~Zhang, L.~Zhou, S.~Liu, Z.~Chen, Y.~Liu, H.~Wang,
  J.~Li, et~al.
\newblock Neural codec language models are zero-shot text to speech
  synthesizers.
\newblock {\em arXiv preprint arXiv:2301.02111}, 2023.

\bibitem{wang17n_interspeech}
Y.~Wang, R.~Skerry-Ryan, D.~Stanton, Y.~Wu, R.~J. Weiss, N.~Jaitly, Z.~Yang,
  Y.~Xiao, Z.~Chen, S.~Bengio, Q.~Le, Y.~Agiomyrgiannakis, R.~Clark, and R.~A.
  Saurous.
\newblock {Tacotron: Towards End-to-End Speech Synthesis}.
\newblock In {\em Proc. Interspeech}, pages 4006--4010, 2017.

\bibitem{wang2018style}
Y.~Wang, D.~Stanton, Y.~Zhang, R.-S. Ryan, E.~Battenberg, J.~Shor, Y.~Xiao,
  Y.~Jia, F.~Ren, and R.~A. Saurous.
\newblock Style tokens: Unsupervised style modeling, control and transfer in
  end-to-end speech synthesis.
\newblock In {\em Proc. Int. Conf. on Mach. Learn.}, pages 5180--5189. PMLR,
  2018.

\bibitem{8413121}
Y.~Xian, C.~H. Lampert, B.~Schiele, and Z.~Akata.
\newblock Zero-shot learning—a comprehensive evaluation of the good, the bad
  and the ugly.
\newblock {\em IEEE Trans. Pattern Anal. Mach. Intell.}, 41(9):2251--2265,
  2019.

\bibitem{10129160}
Y.~Xiao, L.~Wu, J.~Guo, J.~Li, M.~Zhang, T.~Qin, and T.-Y. Liu.
\newblock A survey on non-autoregressive generation for neural machine
  translation and beyond.
\newblock {\em IEEE Trans. Pattern Anal. Mach. Intell.}, 45(10):11407--11427,
  2023.

\bibitem{xue2023multi}
H.~Xue, S.~Guo, P.~Zhu, and M.~Bi.
\newblock Multi-gradspeech: Towards diffusion-based multi-speaker
  text-to-speech using consistent diffusion models.
\newblock {\em arXiv preprint arXiv:2308.10428}, 2023.

\bibitem{yang2023instructtts}
D.~Yang, S.~Liu, R.~Huang, G.~Lei, C.~Weng, H.~Meng, and D.~Yu.
\newblock Instructtts: Modelling expressive tts in discrete latent space with
  natural language style prompt.
\newblock {\em arXiv preprint arXiv:2301.13662}, 2023.

\bibitem{yang2023hifi}
D.~Yang, S.~Liu, R.~Huang, J.~Tian, C.~Weng, and Y.~Zou.
\newblock Hifi-codec: Group-residual vector quantization for high fidelity
  audio codec.
\newblock {\em arXiv preprint arXiv:2305.02765}, 2023.

\bibitem{yang2023uniaudio}
D.~Yang, J.~Tian, X.~Tan, R.~Huang, S.~Liu, X.~Chang, J.~Shi, S.~Zhao, J.~Bian,
  X.~Wu, et~al.
\newblock Uniaudio: An audio foundation model toward universal audio
  generation.
\newblock {\em arXiv preprint arXiv:2310.00704}, 2023.

\bibitem{ye2023comospeech}
Z.~Ye, W.~Xue, X.~Tan, J.~Chen, Q.~Liu, and Y.~Guo.
\newblock Comospeech: One-step speech and singing voice synthesis via
  consistency model.
\newblock {\em arXiv preprint arXiv:2305.06908}, 2023.

\bibitem{10095103}
C.-Y. Yu, S.-L. Yeh, G.~Fazekas, and H.~Tang.
\newblock Conditioning and sampling in variational diffusion models for speech
  super-resolution.
\newblock In {\em IEEE Int. Conf. Acoust., Speech, Signal Process.}, pages
  1--5, 2023.

\bibitem{zeghidour2021soundstream}
N.~Zeghidour, A.~Luebs, A.~Omran, J.~Skoglund, and M.~Tagliasacchi.
\newblock Soundstream: An end-to-end neural audio codec.
\newblock {\em IEEE/ACM Trans. Audio, Speech, Lang. Process.}, 30:495--507,
  2021.

\bibitem{zen2019libritts}
H.~Zen, V.~Dang, R.~Clark, Y.~Zhang, R.~J. Weiss, Y.~Jia, Z.~Chen, and Y.~Wu.
\newblock {LibriTTS: A Corpus Derived from LibriSpeech for Text-to-Speech}.
\newblock pages 1526--1530, 2019.

\end{thebibliography}

\end{document}